\documentclass[reprint,
 amsmath,amssymb,
 aps]{revtex4-1}
\usepackage{graphicx}% Include figure files
\usepackage{dcolumn}% Align table columns on decimal point
\usepackage{bm}% bold math
\def\runauthor{Nicol\`o Cartiglia}
\def\shorttitle{proton-proton total cross section}
\usepackage{wrapfig}
\newcommand{\rs}{\mbox{${\rm \sqrt{s} \;}$}}
 
\newcommand{\stot}{\mbox{${ \sigma_{Tot}(pp)}$ }} 
\newcommand{\ssd}{\mbox{${ \sigma_{SD}(pp)}$ }} 
\newcommand{\sdd}{\mbox{${ \sigma_{DD}(pp)}$ }} 
\newcommand{\sinel}{\mbox{${ \sigma_{Inel}(pp)}$ }} 
\newcommand{\sinelair}{\mbox{${ \sigma_{Inel}(p-air)}$ }} 
\newcommand{\mx}{\mbox{${\rm M_X \;}$}}

 \newcommand {\pomer} {\mbox{ ${I\hspace{-0.25em}P}$ }} 
 \newcommand {\pom}  {I\hspace{-0.15em}P}

\newcommand{\bi} {\begin{itemize}} \newcommand{\ei} {\end{itemize}}
\newcommand{\be} {\begin{equation}} \newcommand{\ee} {\end{equation}}
\newcommand{\bc} {\begin{center}} \newcommand{\ec} {\end{center}}
\newcommand{\noi} {\noindent}

\newcommand{\bfm}{\begin{figure}[htb]\vspace{-0.6cm}}

\begin{document}

\title{Measurement of the proton-proton total, elastic, inelastic and diffractive cross sections  at $\sqrt{s}$ = 2, 7, 8 and 57 TeV}

\author{Nicol\`o Cartiglia\\
(on behalf of the ALICE, ATLAS, CMS and TOTEM collaborations)
}

\affiliation{ INFN \\
Via Pietro Giuria 1, 10125 Torino, Italia }

\begin{abstract}
\noi The measurement of the total $pp$ cross section and its various sub-components (elastic, inelastic and diffractive) is a very powerful tool to understand the proton macro structure and fundamental QCD dynamics. In this  contribution I first provide a theoretical introduction to the topic, then a summary of the experimental techniques and finally I review the new results from AUGER and LHC experiments.

\end{abstract}

\maketitle

\section{Setting the stage} 
\begin{figure}[h]
\begin{center}
 \resizebox{8cm}{!}{\includegraphics{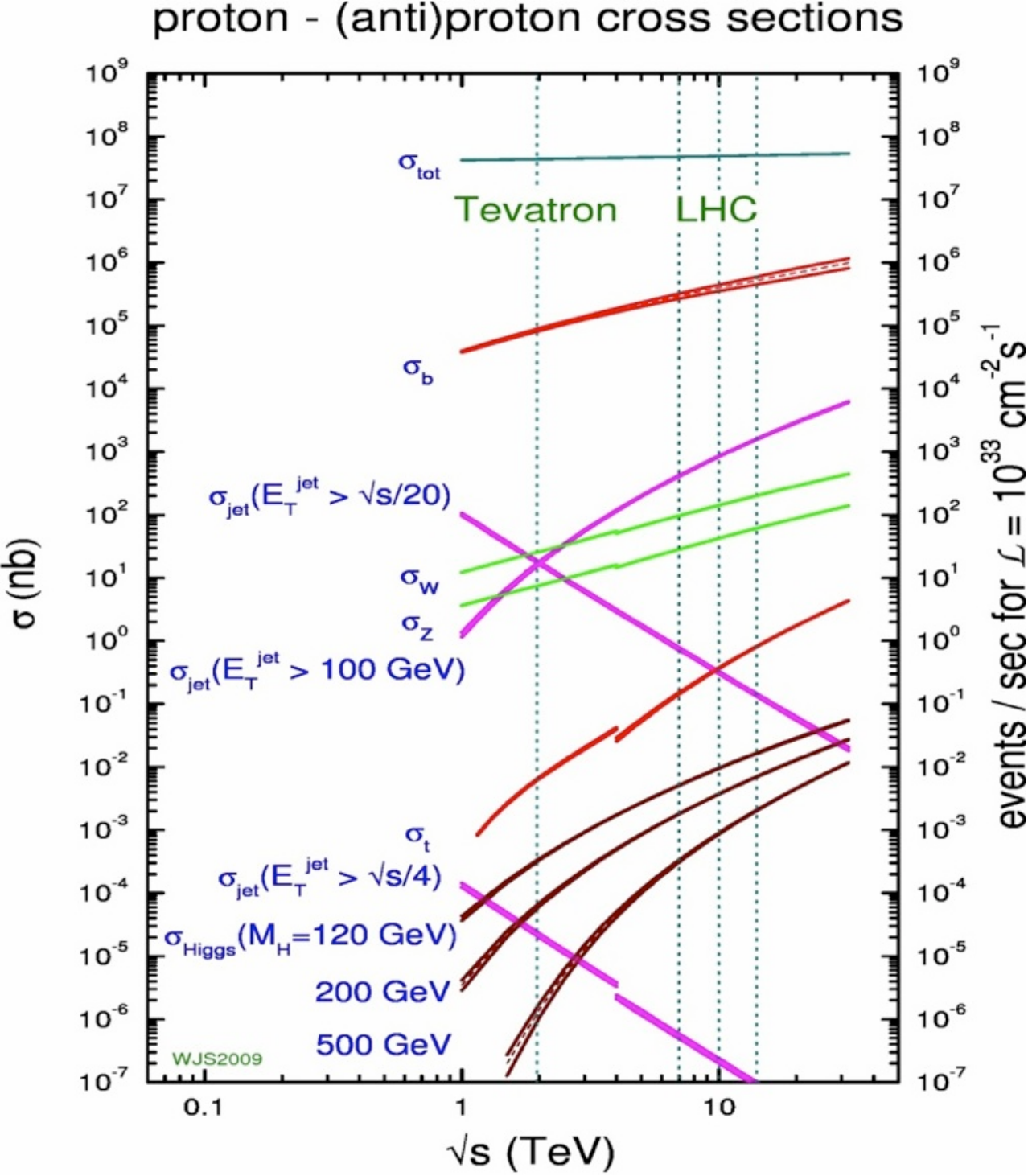}}
\end{center}
 \caption{Cross section values for several important processes. The right vertical axis reports the number of events for a  luminosity value $ L = 10^{33} cm^{-2} s^{-1}$ }
\label{fig:LHCall}
\end{figure}

\noindent Figure~\ref{fig:LHCall} shows the cross section values for many important processes as a function of center-of-mass energy \rs. 

\noi It is clear from the figure  that the value of the total cross section ($\sigma_{Tot}(pp)$) is orders of magnitude larger than  very abundant  QCD processes such as the production of b-quarks. The reason for this fact is that the total cross section is dominated by soft QCD physic processes. The typical value of the total $pp$ cross section is
 $$\sigma_{Tot}(pp) \sim 100 \; mb  \;\; (10^{-25} cm^2),$$ 
which is equivalent to the scattering of two hard balls with a radius of one Fermi ($10^{-13} cm$) each. The geometrical interpretation is really not that useful in particle physics, as the scattering amplitude is often governed by resonance effects which greatly enhance the probability of a process. Consider for example the boron cross section to capture a neutron: it reaches 1200 barns, while the boron geometrical size is about 0.1 barn. The intuitive picture of $pp$ scattering  is further complicated when we consider that the proton is actually a composed objects,  made of valence quarks, $uud$, plus sea quarks and gluons. The valence quarks are what identify a proton as a proton, while the sea-quarks and the gluons can be considered $SU(3)$ symmetric and therefore identical in protons and anti-protons (this last statement is not entirely true, but the non $SU(3)$ symmetric part does not have an impact on the measurement of the total cross section). \\

\begin{figure}[h]
  \begin{center}
\resizebox{9cm}{!}{\includegraphics{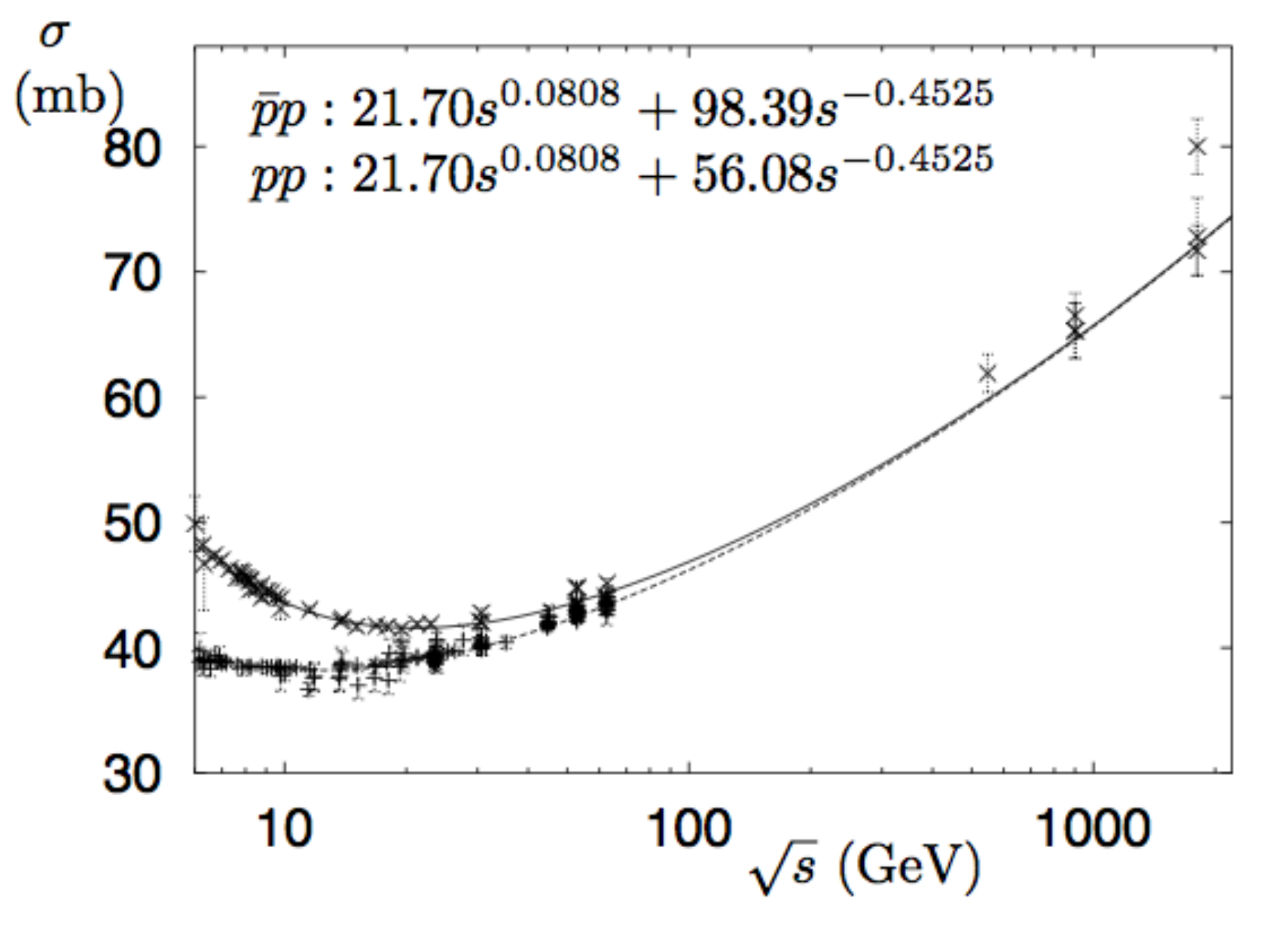}}
  \end{center}
    \caption{Values of the $pp$ and $p\bar{p}$ total cross section as a function of  \rs (\cite{Pom}) }
     \label{fig:low}
\end{figure}

\noi Figure~\ref{fig:low}  shows the value of  the $pp$ and $p\bar{p}$ total cross section as a function of  \rs, up to 2 TeV. From this figure we can already understand a very important fact: at low  energy, $\sqrt{s} \le $ 200 GeV, the values of $\sigma_{Tot}(pp)$ and  $\sigma_{Tot}(p\bar{p})$ are different and therefore the valence quarks must play an important role, whilst at higher energy the values of $\sigma_{Tot}(pp)$ and  $\sigma_{Tot}(p\bar{p})$ are the same, indicating that the scattering is dominated by the $SU(3)$-symmetric component.

\section{Theoretical framework} 

\begin{figure}[h]
  \begin{center}
\resizebox{9cm}{!}{\includegraphics{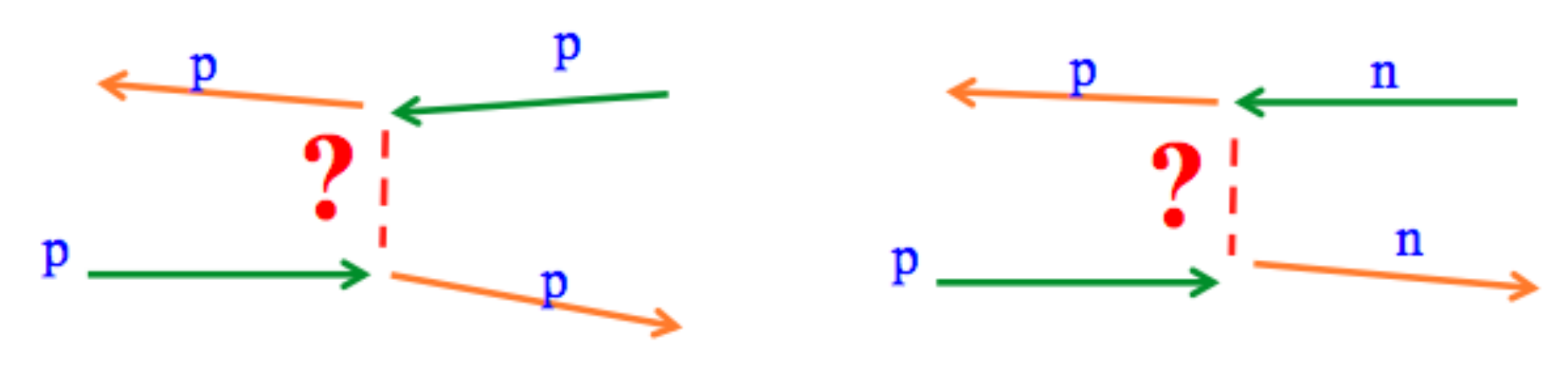}}
  \end{center}
    \caption{Schematic of the elastic $pp \rightarrow pp$ and charge-exchange $pn\rightarrow np$ channel.}
     \label{fig:regge1}
\end{figure}

\noi A large part of the total cross section is due to  soft processes such as the elastic channel, Figure~\ref{fig:regge1} (left pane) or charge-exchange reactions,  Figure~\ref{fig:regge1} (right pane).  It is therefore important to understand what is the mechanism underlying these processes and what  theoretical models can be used to quantify them.

\begin{figure}[h]
\begin{center}
 \resizebox{8cm}{!}{\includegraphics{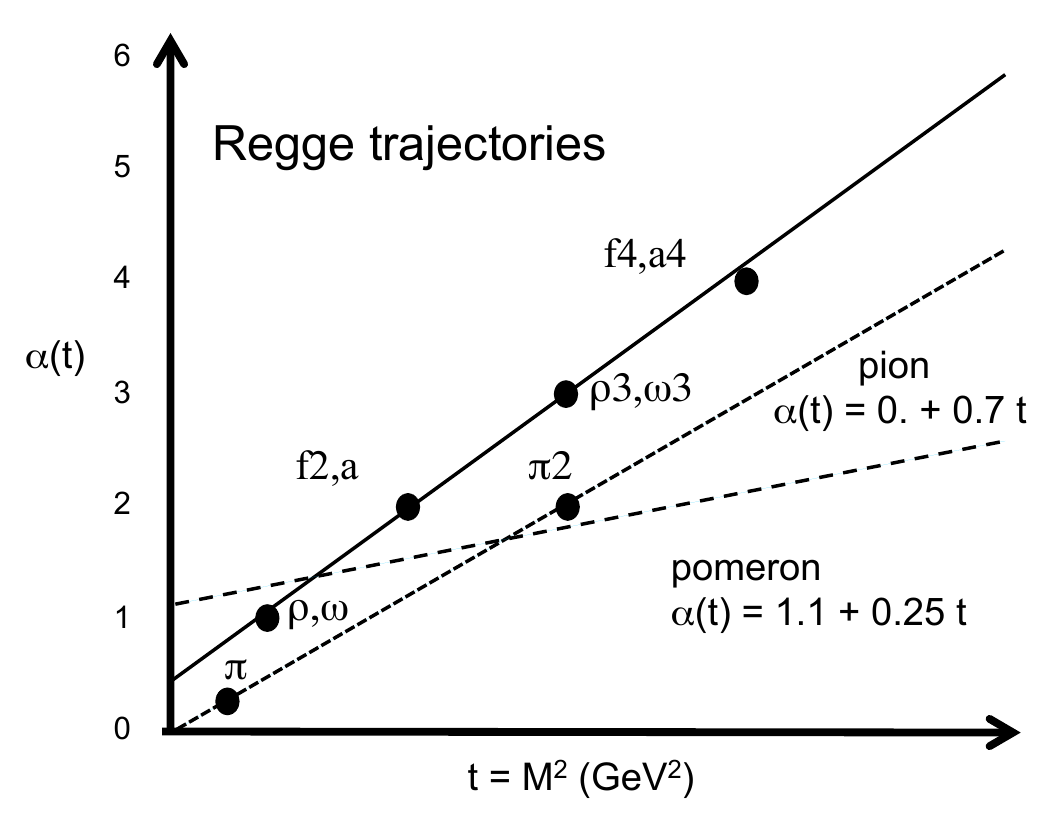}}
\end{center}
\caption{Example of Regge trajectories in the complex angular moment (J) - t plane:  for values of $t = m^2$ (where $m$ is the mass of a particle in the trajectory)  $\alpha(t)$ corresponds to the spin of the particle.}
\label{fig:regge_N}
\end{figure}

\noi Regge theory~\cite{Regge:first} is the  framework that is used to study many soft QCD processes such as diffraction and the total cross section and it describes high-energy scattering at small $t$ in terms of the exchanges of mesons and possibly glueballs. For a detailed discussion about pomeron physics and QCD see for example~\cite{Pom}. The first main feature of Regge theory is based on  the observation that group of particles order themselves in straight lines (trajectories) when plotted  in the complex angular moment $J$ - $t$ plane, where $t$ is  the 4-momentum transfer squared.  
$$\alpha(t) = \alpha + \alpha\prime t.$$ 
The particles are such that, whenever $t = m^2$ (where $m$ is the mass of a particle in the trajectory), then $\alpha(t)$ correspond to the spin of the particle, Figure~\ref{fig:regge_N}.

\noindent Regge theory explains that these trajectories can be understood as group of particles that are exchanged together, i.e., referring to Figure~\ref{fig:regge_N}, in the scattering process  $p n \rightarrow n p $ not just the $\pi$ particle is exchanged, but all particles on the $\pi$ trajectory. Mathematically, each particle is a {\it pole} in the analytic expression of the scattering amplitude of processes mediated by its own trajectory. The second main feature of the Regge pole model is the relationship between exchanged trajectories and high-energy behaviour:  a given trajectory contributes to  \stot according to:
$$\sigma(s)  \propto Im A(s, t = 0) \sim s^{\alpha-1},$$  
where $Im A(s, t = 0)$ is the imaginary part of the scattering amplitude computed at $ t = 0 $ GeV and  $\alpha$ is the intercept of the exchanged trajectory. The equation of the trajectory indicates a very important feature: if the intercept is lower than one, the contribution of a trajectory to \stot decreases as a function of increasing \rs. \\

\begin{figure}[h]
\begin{center}
 \resizebox{8cm}{!}{\includegraphics{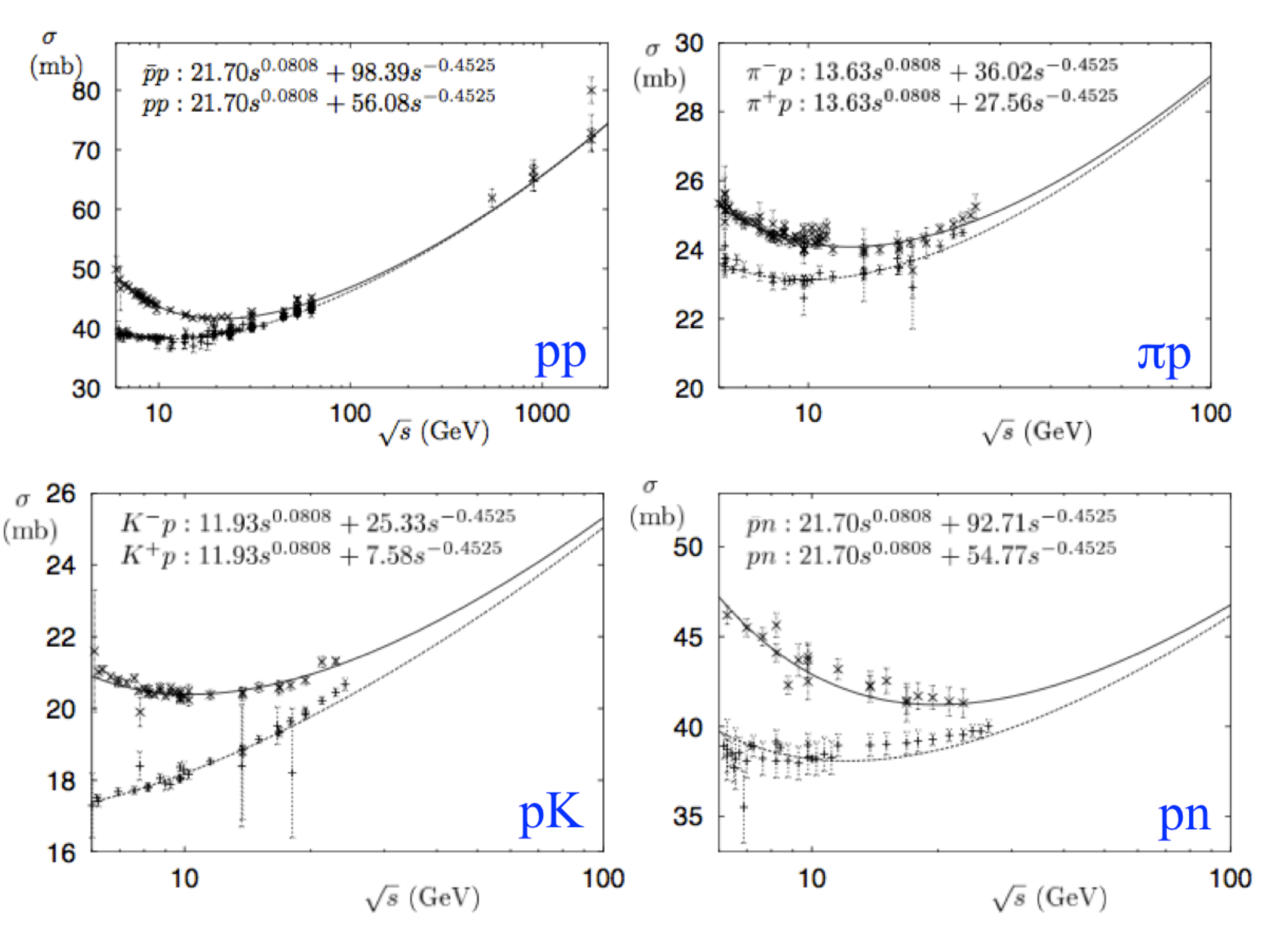}}
\end{center}
\caption{ Behaviour of the total cross section with the center of mass energy for several different scattering processes $pp, p\bar{p}, \pi^+ p, \pi^-p, K^- p, K^+p, pn, \bar{p}n$ (\cite{Pom}). }
\label{fig:cross_sections}
\end{figure}

\noi An interesting fact happens: each known  trajectory has the intercept lower than one and therefore provides a decreasing  contribution to \stot. This prediction is, however, not supported by the experimental points: following an initial decrease of the cross section with \rs that follows the behaviour predicted by the exchange of reggeon trajectories ($\sigma(s)  \propto  \sim s^{-0.5}$),  the value of the cross section rises  with the  trend $\sigma(s)  \propto  \sim s^{0.08}$. This general feature is displayed  in Figure~\ref{fig:cross_sections} where the value of \stot  versus \rs is plotted for several different scattering processes ($pp, p\bar{p}, \pi^+ p, \pi^-p, K^- p, K^+p, pn, \bar{p}n$).

\noi This contradiction  is eliminated by introducing a new trajectory with an intercept slightly larger than one: the pomeron trajectory, also shown on Figure~\ref{fig:regge_N}. This special trajectory does not have any on-shell particle on it, and therefore it could not have been measured using the known particles. The pomeron trajectory is probably related to the exchange of glueballs, however the experimental measurements of glueball states up to now do not support or deny this idea. The intercept of the pomeron trajectory is traditionally written as:
$$ \alpha_{Pom}(t=0) = 1+ \Delta$$
with, according to Figure~\ref{fig:cross_sections}, $\Delta = 0.08 $. \\

\noi For a given scattering process,  the exchanged particles (poles) on the reggeon and pion trajectories offer guidance on how to write the scattering amplitude $A(s, t)$,  however this is not the case for the pomeron trajectory, as it has no particles on it: the analytical form of $A(s, t)$ for pomeron exchange is less constrained and it depends on  the type of diagram considered. The possible contributions of the pomeron trajectory to the total cross section can contain different terms: 

\begin{eqnarray}
\label{eq:pom1}
\sigma(s)  \propto Im A(s, t = 0) \sim s^{\alpha-1}, \\
\label{eq:pom2}
\sigma(s)  \propto Im A(s, t = 0) \sim ln(s), \\
\label{eq:pom3}
\sigma(s)  \propto Im A(s, t = 0) \sim ln^2(s), 
\end{eqnarray}

\noindent where Equation~\ref{eq:pom1} is for a simple poles type of exchange, while Equation~\ref{eq:pom2} and Equation~\ref{eq:pom3} are for more complicate processes. \\

\begin{figure}[h]
\begin{center}
 \resizebox{8cm}{!}{\includegraphics{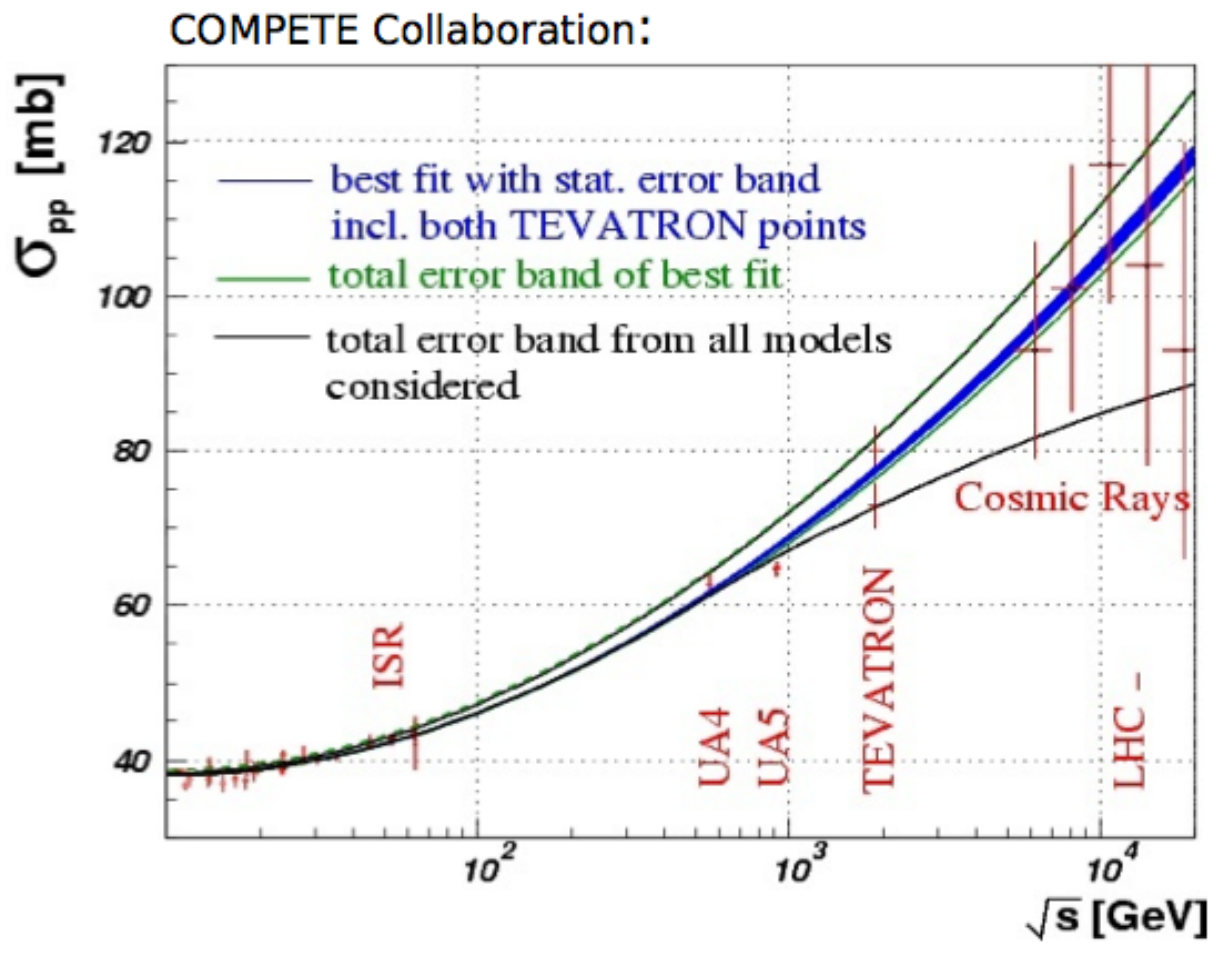}}
\end{center}
\caption{ Evolution of the value of total $pp$ cross section as a function of  \rs  as predicted by the COMPETE collaboration. The darkest band is the fit that has the best $\chi^2/DOF$ using pre-LHC points.}
\label{fig:compete}
\end{figure}

\noi The three most common parametrizations of the cross section are:
\begin{eqnarray}
\label{eq:cross1}
\sigma(s)  =  c_1 + c_2 * s^{-0.5} + c_3 * s^{0.08}, \\
\label{eq:cross2}
\sigma(s)  =  c_1 + c_2 * s^{-0.5} + c_3 * ln^2(s), \\
\label{eq:cross3}
\sigma(s)  =  c_1 + c_2 * ln(s) + c_3 * ln^2(s). 
\end{eqnarray}
\noi The various diagrams have been analyzed by the COMPETE~\cite{Nicolescu:2001um}\cite{Cudell:2001ma} collaboration, which has produced a prediction for the evolution of the  value of the total $pp$ cross section as a function of \rs, Figure~\ref{fig:compete}. These studies find that the analytic form that fits the low energy data points better  is Equation~\ref{eq:cross2}.

\noi Their best  {\it pre-LHC} predictions are: 

\begin{eqnarray}
\sigma_{Tot}(7 \; TeV) = 98 \pm 5 \;\; mb, \\
\sigma_{Tot}(8\; TeV) = 101 \pm 5 \;\; mb, \\  
\sigma_{Tot}(14 \; TeV) = 111.5 \pm 5\;\; mb. 
\end{eqnarray}

\subsection{The rise of the gluon distribution}

\noindent Following the experimental discovery at HERA of the steep rise of the gluon distribution as a function of \rs, the predictions for the value of the total cross section at LHC energy were  updated. In particular, Equation~\ref{eq:cross1} was modified introducing a second simple pole, the so called {\it hard pomeron}~\cite{Landshoff:2007uk} \cite{Cudell:2009bx}: 

\begin{equation}
\label{eq:hard}
\sigma(s)  =  c_1 + c_2 * s^{-0.5} + c_3 * s^{0.067} + c_4 * s^{0.45}.
\end{equation}

\begin{figure}[h]
\begin{center}
 \resizebox{8cm}{!}{\includegraphics{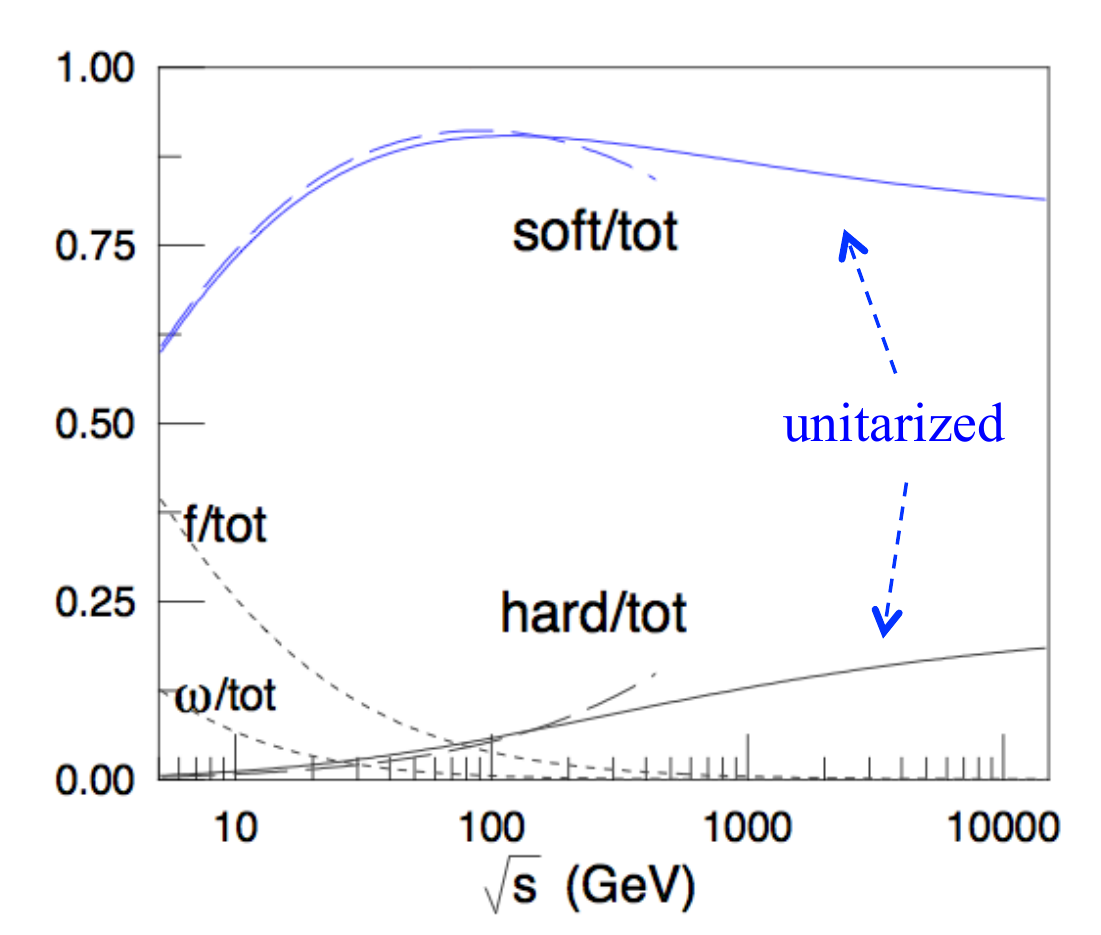}}
\end{center}
\caption{Various contributions to the cross section as a function of  \rs in the two-pole parametrization (\cite{Cudell:2009bx}).}
\label{fig:softhard}
\end{figure}

\noindent The combined fit using the {\it soft} and {\it hard} pomerons lowers the soft pomeron intercept from $0.08$ to $0.067$ while the hard pomeron intercept value of $0.45$ is perfectly compatible with HERA results. Figure~\ref{fig:softhard}  shows the various contributions to the cross section as a function of \rs.

\subsection{Paradoxes, bounds and unitarization effects}

\noi The Regge  formalism outlined above  leads to predictions that cannot be accurate at very high energies. For example, it is clear that the cross section cannot rise indefinitely as $\sigma \propto s^{\Delta}$. \\

\noi Using s-channel unitarity and S-matrix analyticity constraints it is possible to derive the so called Froissart-Martin bound:
$$ \sigma_{Tot}(pp) \le C*ln^2(s/s_o) $$
with $s_o \sim 1$ GeV and $C = \pi / 2m_\pi^2 $ = 30 mb. Even though this bound has no effect at LHC energies as it requires $\sigma(7\; TeV)_{Tot} < 2.3$ barn and  $\sigma(14\; TeV)_{Tot} < 2.7$ barn, it puts a limit on the possible growth of \stot.\\

\noi Likewise, the Regge formalism  predicts that the elastic cross section grows with energy faster than the total cross section
$$\sigma_{El} \propto s^{2\Delta}, $$
creating the paradox that at a certain energy the elastic cross section would be larger than the total cross section. This condition, however, does not happen as  it can be shown that the elastic cross section should always be less than half of the total cross section: $\sigma_{El} < 1/2 \sigma_{Tot} $, the so called Pumplin bound.

\begin{figure}[h]
\begin{center}
 \resizebox{8cm}{!}{\includegraphics{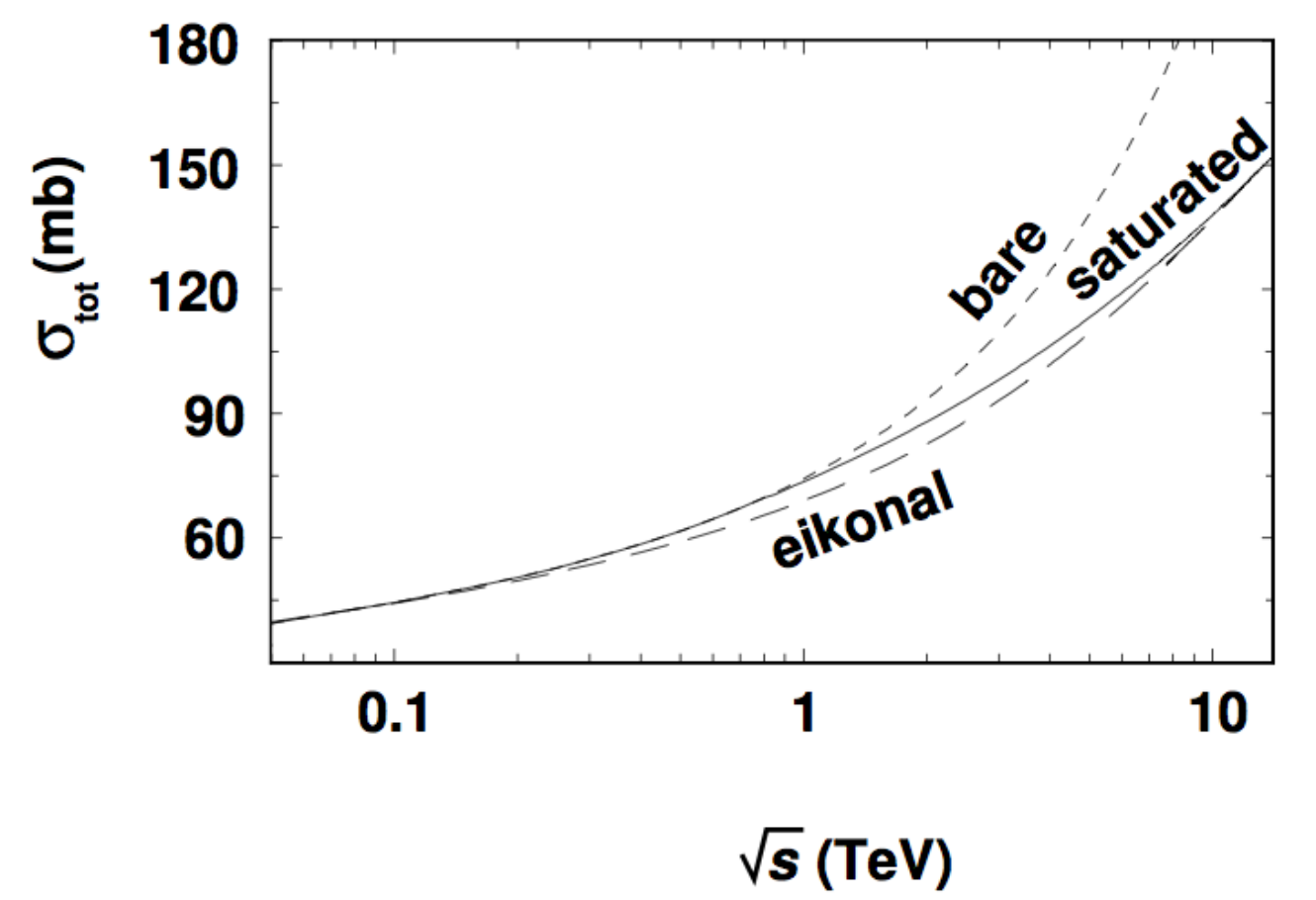}}
\end{center}
\caption{Predictions of the two-pomeron model without ({\it bare}) and with ({\it eikonal, saturated}) unitarization effects (\cite{Cudell:2009bx}). }
\label{fig:unitar}
\end{figure}

\noi The inclusion of these bounds and the effects of multiple exchanges in the calculation of  the value of the total cross section is called  {\it unitarization}. The overall effect of unitarization is to reduce the value of the total cross section: Figure~\ref{fig:unitar} shows the predictions of the two-pomeron model without ({\it bare}) and with ({\it eikonal, saturated}) unitarization effects. It's interesting to note that the prediction from a given model is the outcome of the interplay of its functional expression and the unitarization scheme used.  \\

\section{Proton - proton elastic scattering} 
\begin{figure}[h]
\begin{center}
 \resizebox{8cm}{!}{\includegraphics{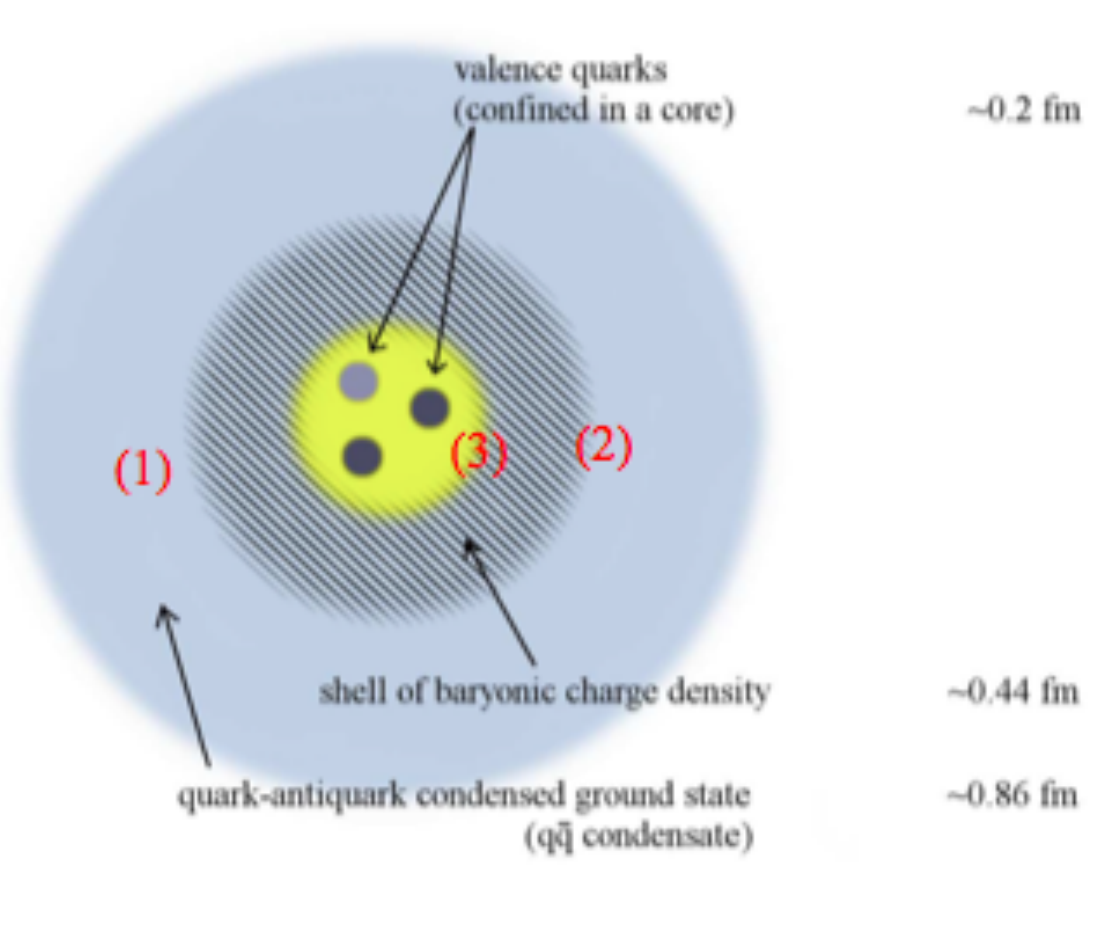}}
\end{center}
\caption{Sketch of the proton macro structure (see text for details) (\cite{Islam}).}
\label{fig:pshells}
\end{figure}

\noindent Elastic scattering, $pp \rightarrow pp$, is a very important process to probe the macro structure of the proton, and it represents roughly one forth of the total cross section.  A sketch of the proton macro structure, following~\cite{Islam}, is shown in Figure~\ref{fig:pshells}: the outer corona (1) is composed by $q\bar{q}$ condensates, the middle part is a shell of baryonic charge density (2), while  the valence quarks are confined at the center (3).  

\begin{figure}[h]
\begin{center}
 \resizebox{6cm}{!}{\includegraphics{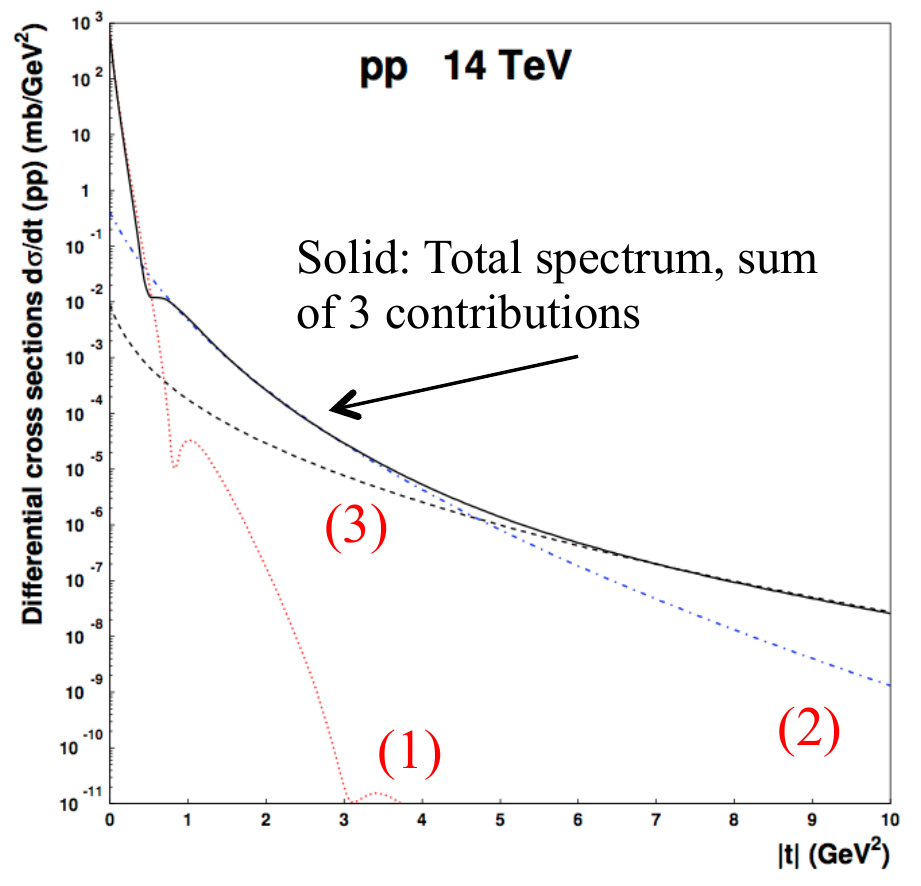}}
\end{center}
\caption{Elastic cross section as a function of the 4-momentum transfer $t$. The contributions from the various layers of the proton are shown  (\cite{Islam}).}
\label{fig:pstruct}
\end{figure}

\noindent Elastic scattering probes the proton at a distance $b$ given by $ b \sim 1/\sqrt{t}$. At low $t$ values, the cross section is well approximated by an exponential  form: 
$$ \frac{d\sigma}{dt} = A * e^{Bt}, $$ 
and is largely dominated by the outer corona (1), Figure~\ref{fig:pstruct}.

\noi At higher values of $t$, the cross section has a more complex form, reflecting the additional contributions from inner layers. At values of $t$ above 4 GeV$^2$ the cross section is dominated by quark-quark elastic scattering (the so called {\it deep elastic scattering}). \\

\begin{figure}[h]
\begin{center}
 \resizebox{6cm}{!}{\includegraphics{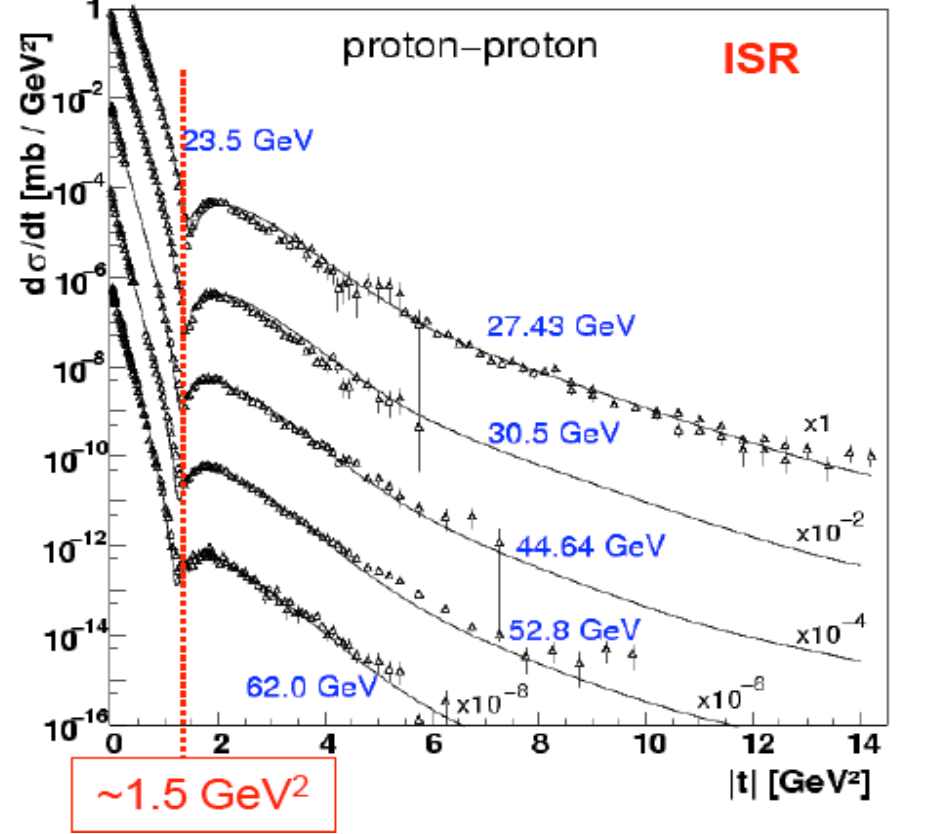}}
\end{center}
    \caption{Differential value of the proton-proton cross section as a function of the 4-momentum transfer squared $t$, for different center-of-mass energies.}
     \label{fig:isr}
\end{figure}

\noi As  \rs grows, not only the proton becomes blacker (the cross section increases) but also grows in size: the value of the slope parameter $B$ increases (the so called ``shrinkage of the forward peak''), indicating an average lower values of $t$ and therefore a longer interaction range. The relative importance of the various components described in Figure~\ref{fig:pstruct}  changes with energy as shown by the experimental data from ISR, Figure~\ref{fig:isr}, where the  values of the position and depth of the dip decrease with increasing \rs.

\section{Montecarlo models}

\noi The Montecarlo (MC) models commonly used in high-energy and cosmic-rays physics can be approximately divided into two large families. The  MC models in the first group (QGSJET, SIBYLL, PHOJET, EPOS) are based on Regge Field Theory (RFT), and they differ among themselves for the implementation of various aspects of the model parameters. For their focus on soft and forward physics they have been intensively used in cosmic-rays physics and they are a very important tool in the study of the total cross section. These MC models have been extended to also provide predictions for  hard QCD processes. \\

\noi The second group of MC models is based on the calculation of perturbative QCD matrix elements (PYTHIA, HERWIG, SHERPA) and the relative importance and absolute values of  soft processes (total and inelastic cross section, fraction of non-diffractive and  diffractive events) just reflects the chosen internal parametrization. \\

\begin{figure}[h]
\begin{center}
 \resizebox{8cm}{!}{\includegraphics{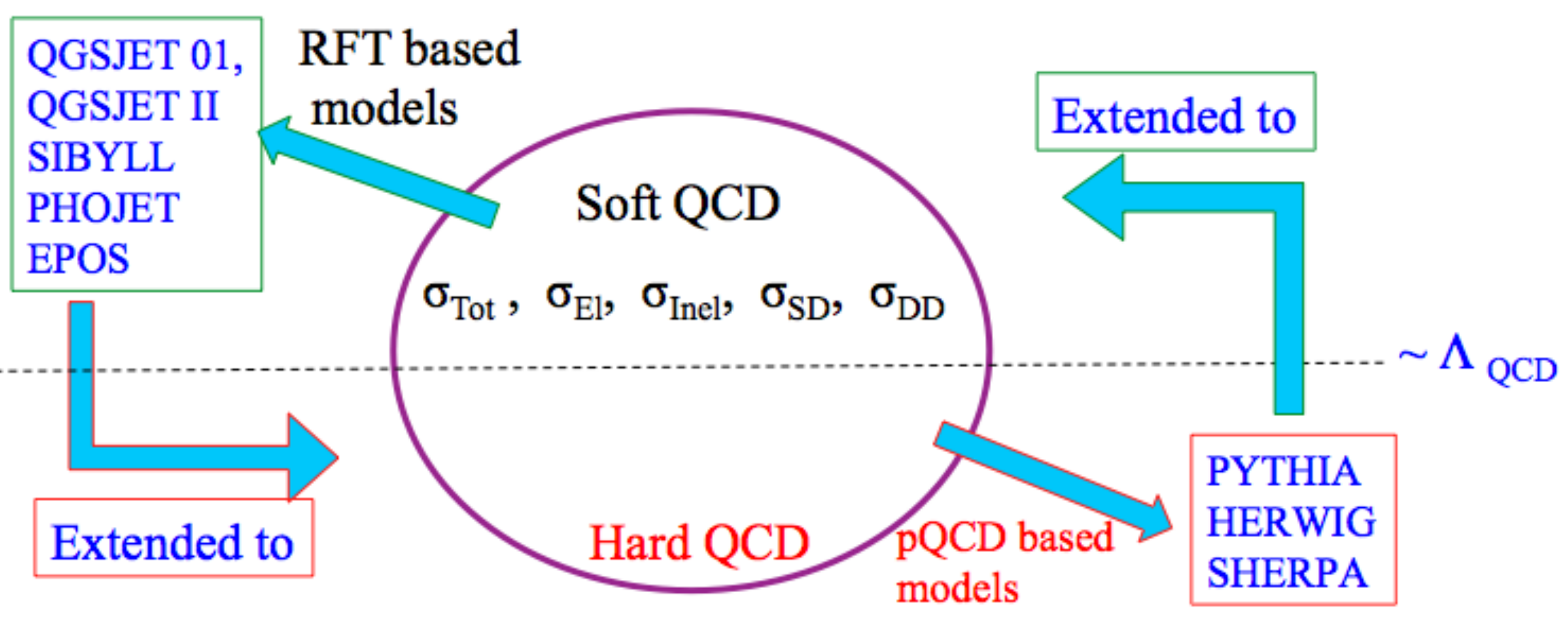}}
\end{center}
\caption{Montecarlo models commonly used in high-energy and cosmic-rays physics}
\label{fig:mc}
\end{figure}

\noi Both families can be used when studying soft QCD, but these differences, outlined in  Figure~\ref{fig:mc}, need to be considered when interpreting the results.

\section{Topologies of events in $\sigma_{Tot}(pp)$ }

%\begin{wrapfigure}{r}{0.5\textwidth}
%  \begin{center}
%    \includegraphics[width=0.48\textwidth]{components.pdf}
%  \end{center}
%    \caption{Differential value of the proton-proton cross section as a function of the 4-momentum transferred, for different center-of-mass energies.}
%     \label{fig:comp}
%\end{wrapfigure}

Three main components contribute to the value of $\sigma_{Tot}(pp)$.
\begin{itemize}
\item [(i)] Elastic scattering $pp \rightarrow pp$: 20-25\% of $\sigma_{Tot} (pp)$. \\
\noi This is a very difficult process to measure and requires dedicated hardware, often installed and operated by dedicated experiments. 
\item [(ii)] Diffractive scattering $pp \rightarrow Xp; \; pp \rightarrow pXp; \; pp \rightarrow XY$: 25 - 35 \% of $\sigma_{Tot}(pp)$. \\
\noi This group includes all processes that are mediated by the exchange of a {\it pomeron}, \pomer. These events are recognizable by the presence of a large gap in the rapidity distributions of final state particles. For  Reggeon or Pomeron exchange the probability to have a 
rapidity  gap $\Delta\eta$ depends on the value of the intercept of the exchanged trajectory\cite{Kopeliovich:1996iw}:
$${\rm p(\Delta\eta) \sim e^{-2 ( \alpha(0) -1)\Delta\eta} .}$$
Let's then consider different possibilities:
\begin{itemize}
\item \pomer: ${\rm \alpha_{\pom}(0) \sim 1 \;\; \Rightarrow  \;\; p(\Delta\eta) \sim e^{0}}$ 
\item ${\rm \rho,a_2, f_2,\omega}$: ${\rm \alpha_R(0) \sim 0.5 \; \; \Rightarrow \;\; p(\Delta\eta) \sim e^{-\Delta\eta}} $ 
\item ${\rm \pi}$: ${\rm \alpha_{\pi}(0) \sim 0 \; \; \Rightarrow \;\; p(\Delta\eta) \sim e^{-2\Delta\eta}}. $ 
\end{itemize}

\noi Therefore, even though  ${\rm \rho, \pi \; and \; \pomer}$ are colourless exchanges,
only \pomer exchange produces rapidity gaps that are not suppressed as the gap width increases. It is therefore possible to operationally define  diffraction~\cite{ref:bj} by the presence of a rapidity gap: diffractive events are those which lead to a large rapidity gap in final state phase space and are not exponentially suppressed as a function of the gap width.

\noi This process includes single, double and central diffraction. Low mass diffraction is experimentally  difficult to measure as the hadronic activity produced in the interaction is very small. 
\item [(iii)] Non-diffractive scattering (everything else): 50-60\%.\\
\noi This is the easiest part to measure, as normally the events have a large number of particles and can be easily detected. 
\end{itemize}

\noi The distinction between elastic and non-elastic scattering is quite obvious, therefore often the results are presented in terms of three components: $\sigma_{Tot}(pp), \; \sigma_{El}(pp)$ and $\sigma_{Inel}(pp)$. On the other hand,  the distinction between diffractive and non-diffractive events is much less straightforward and it requires MC models to asses the efficiency of detecting rapidity gaps in the distribution of final state particles. The values of the single, \ssd, and double, \sdd, diffraction cross sections are therefore more difficult to evaluate, and they are often quoted inside certain limits, either in mass or rapidity.

\section{Measurement of $\sigma_{Tot}(pp), \;  \sigma_{El}(pp)$, and $\sigma_{Inel }(pp) $ using elastic scattering.}

\begin{figure}[h]
\begin{center}
 \resizebox{8cm}{!}{\includegraphics{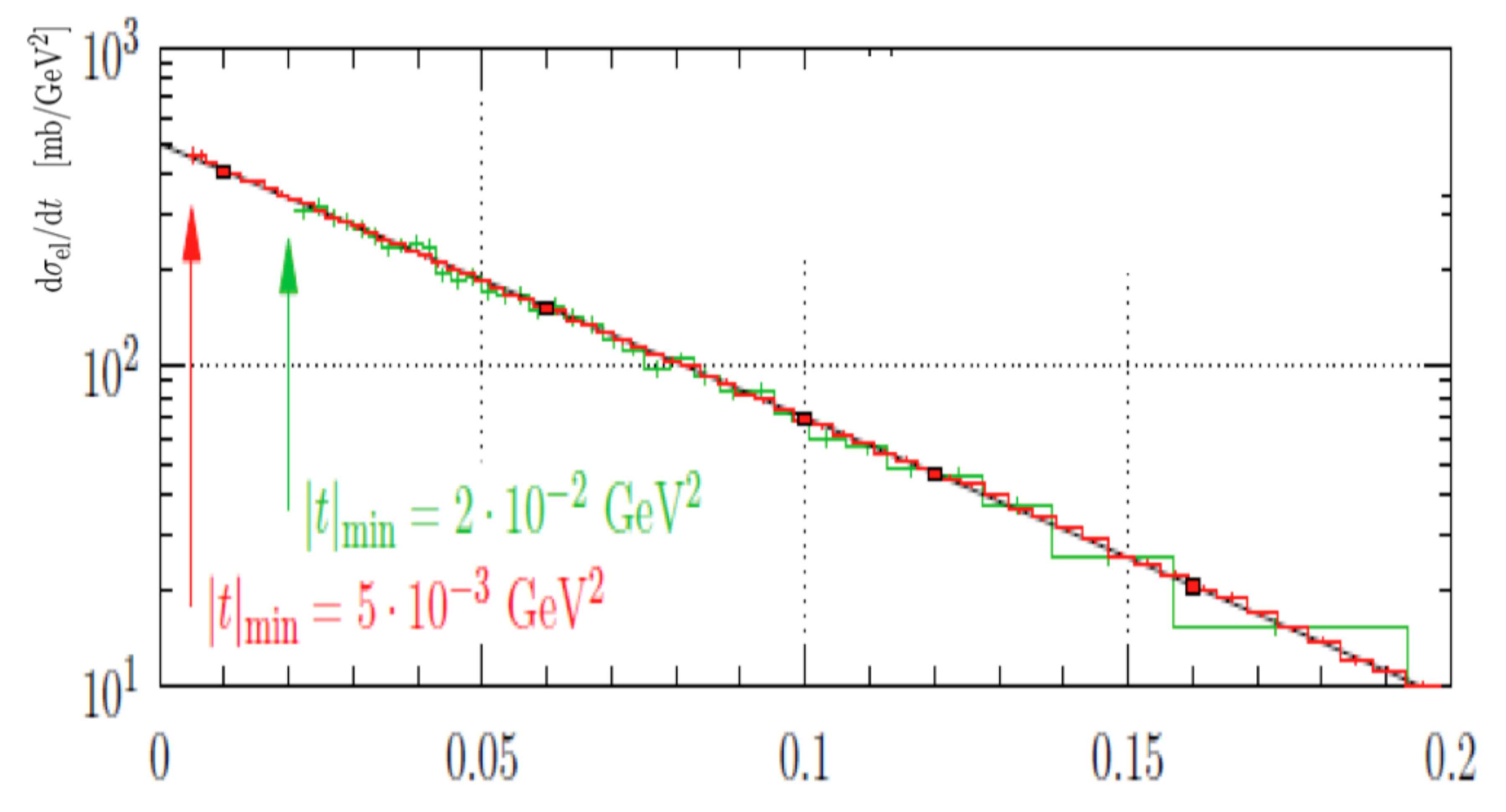}}
\end{center}
\caption{ $d\sigma/dt$ elastic differential $pp$ cross-section in the very low $t$ range, showing the extrapolation to $t = 0 $ GeV/c.}
\label{fig:dt}
\end{figure}

\noindent The TOTEM experiment at the LHC has the capability to detect elastic $pp$ scattering and to measure the differential cross-section for elastic $pp$ scattering as a function of the four-momentum transfer squared $t$ \cite{Antchev:2013gaa}. The differential $d\sigma/dt$ distribution is shown in Figure~\ref{fig:dt}. 

\noi The elastic data were integrated up to $|t| = 0.415 $ GeV$^2$, where the effect of the larger $|t|$-contributions is small compared to the other uncertainties. The optical theorem can be used to calculate the total $pp$ cross-sections from the value of the $t$ distribution extrapolated to $t = 0$ GeV$^2$:
\begin{equation}
\label{eq:opt}
\sigma^2_{Tot} = \frac{16\pi (\hbar c)^2}{1+\rho^2}\frac{d\sigma_{El}}{dt}|_{t = 0} \;\; \rho = \frac{Re (f_{El})}{Im (f_{El})}|_{t = 0}
\end{equation}
\noindent where the $\rho$ is the ratio of the real to the imaginary part of the forward scattering amplitude. The value of $\rho$ has both been predicted by the COMPETE collaboration ($ \rho^2 \simeq 0.02$) ~\cite{Cudell:2002xe} and measured by TOTEM ($\rho^2 = 0.009 \pm 0.056$)~\cite{Antchev:2013iaa}. In order to derive the elastic and total cross-sections, the extrapolation to the optical point $t = 0$ GeV$^2$ was performed.  \\

\noi The minimum $t$ value that can be reached using the TOTEM roman pots depends on the LHC optics: as the angular beam spread $\sigma(\theta)$ is inversely correlated with the accelerator parameter $\beta^*$, $\sigma(\theta)  \propto 1/\sqrt{\beta^*}$, it is necessary to have large $\beta^*$ values to achieve a small beam angular spread and therefore be able to  measure low values of $t$. In 2012 the largest value of  $\beta^*$ achieved by the LHC has been $\beta^*= 1000$ m, allowing to detect scattered protons with  $t \simeq 5* 10^{-4}$ GeV$^2$. 

\begin{figure}[h]
\begin{center}
 \resizebox{7cm}{!}{\includegraphics{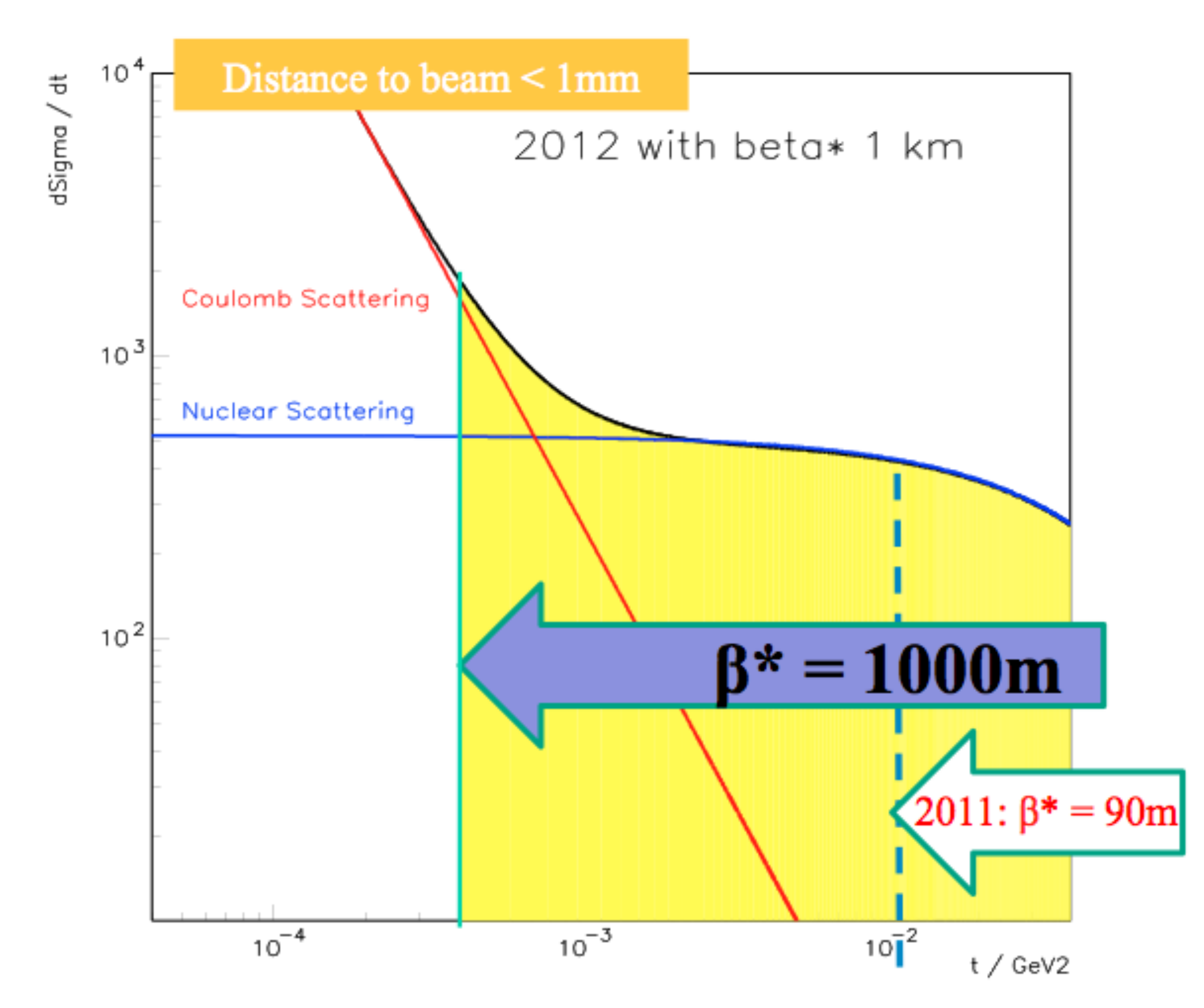}}
\end{center}
 \caption{Reachable part of the  $t$ distribution for two $\beta^*$ values, assuming a distance of the detectors from the beam of less than one millimeter.}
\label{fig:betastar}
\end{figure}

\begin{figure}[h]
\begin{center}
 \resizebox{6cm}{!}{\includegraphics{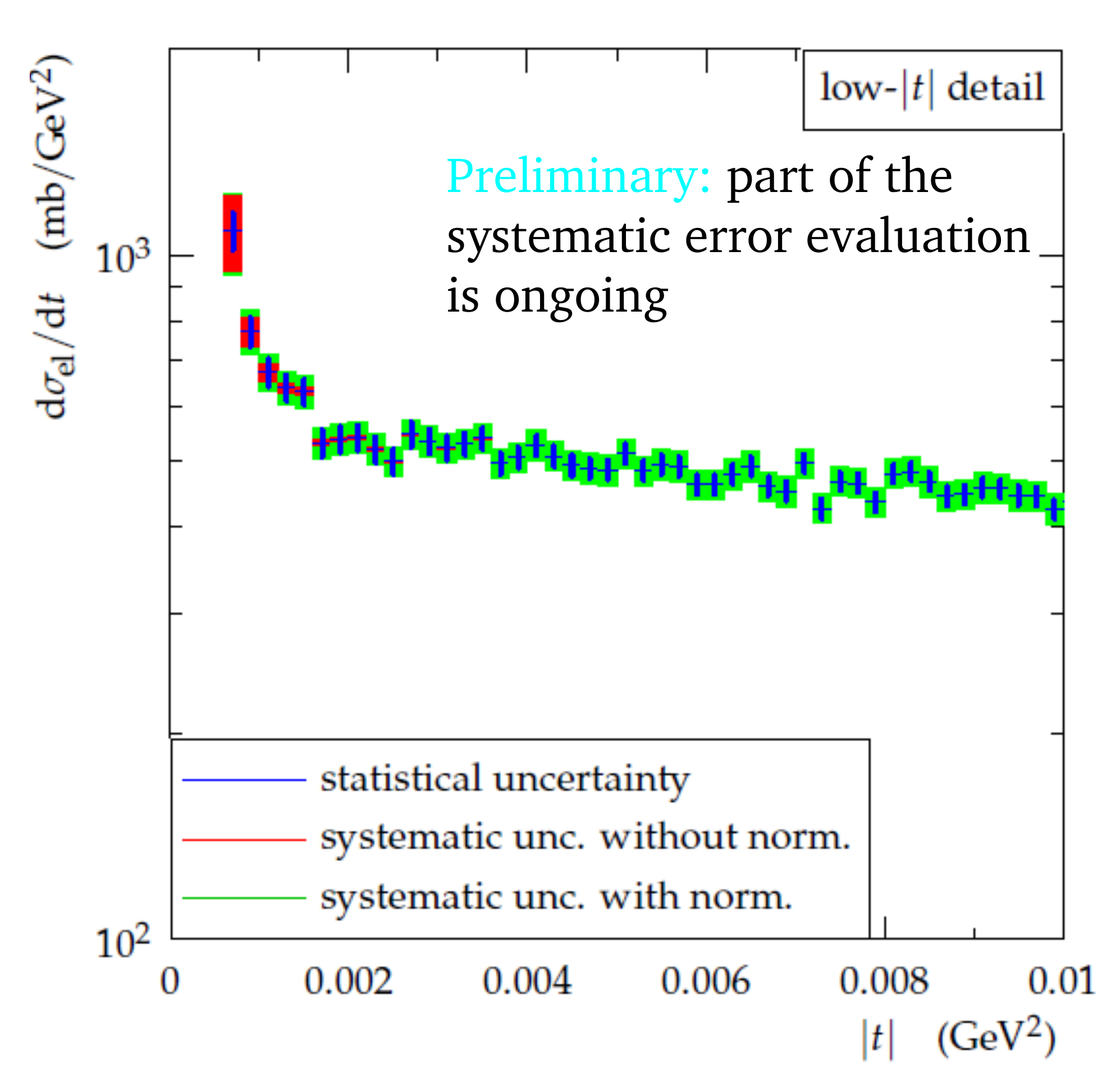}}
\end{center}
\caption{Elastic $pp$ $d\sigma/dt$ distribution at \rs = 8 TeV for very low values of $t$ as measured by the TOTEM collaboration.}
\label{fig:lowt}
\end{figure}

Figure~\ref{fig:betastar} schematically shows the reachable part of the $t$ distribution for two $\beta^*$ values, assuming a distance of the detectors from the beam of  about one millimetre. 
Such a high $\beta^*$ value allowed the TOTEM collaboration to measure the $t$-distribution in the Coulomb region~\cite{DIS2013:totem}, as shown in Figure~\ref{fig:lowt}. This measurement will allow to determined the $\rho$ parameter, validating the model of the  interference between the hadronic and Coulomb amplitudes and reducing the error on the determination of the cross section. \\

\noi In the low $t$ range ($ t < 0.4$ GeV$^2$), the TOTEM collaboration fitted the $d\sigma/dt$ distribution using a simple exponential:
$$ \frac{d\sigma}{dt} \propto e^{Bt},$$ 
and obtained a  value of the slope parameter given by $$ B = 19.9 \pm 0.3 \;\;  {\rm GeV^{-2}}.$$ This value confirms that the {\it shrinkage of the forward peak} continues at LHC energy.  \\

\begin{figure}[h]
\begin{center}
 \resizebox{8cm}{!}{\includegraphics{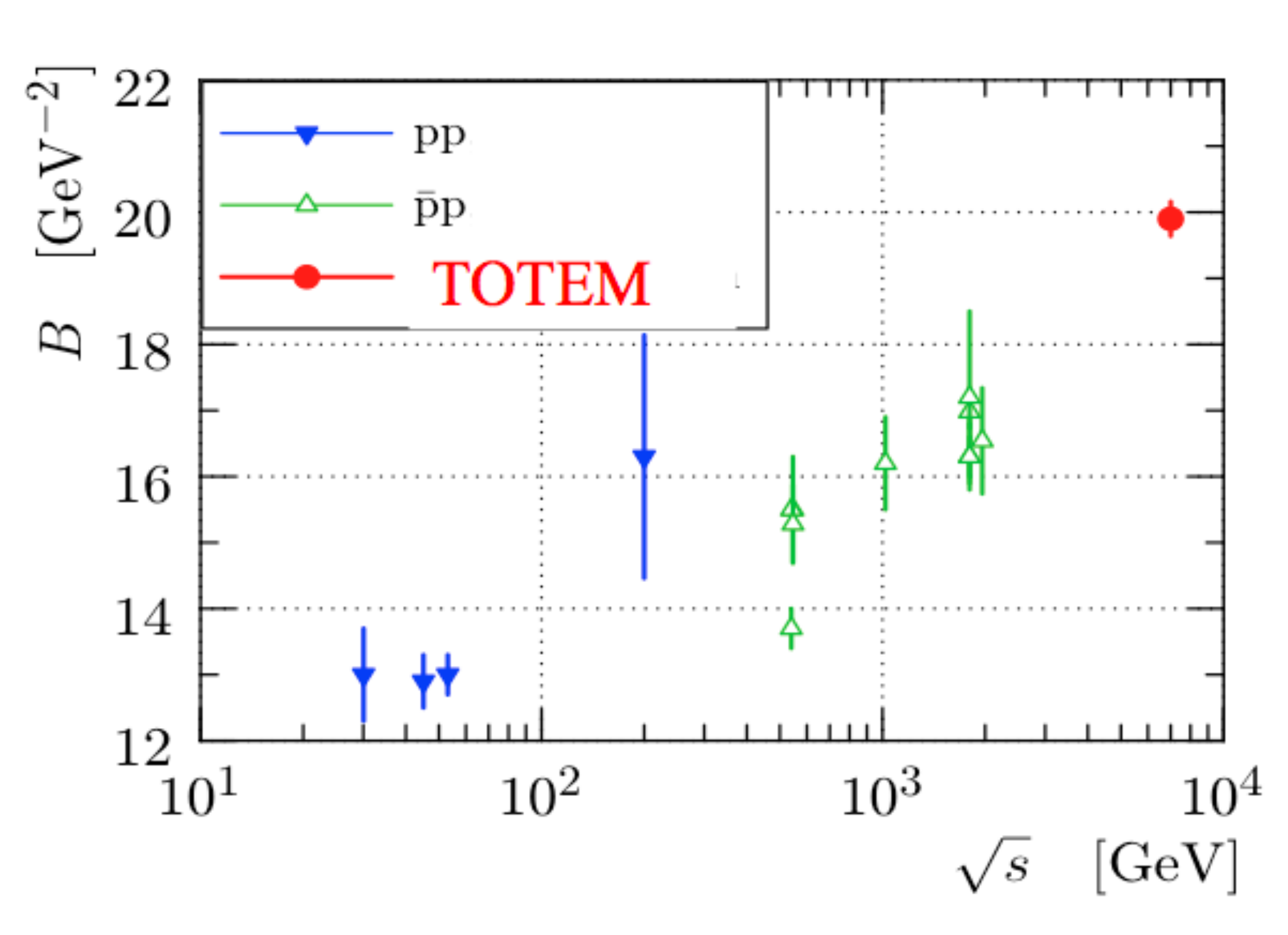}}
\end{center}
 \caption{Value of the elastic slope parameter B as a function of \rs. }
\label{fig:beta}
\end{figure}

\begin{figure}[h]
\begin{center}
 \resizebox{8cm}{!}{\includegraphics{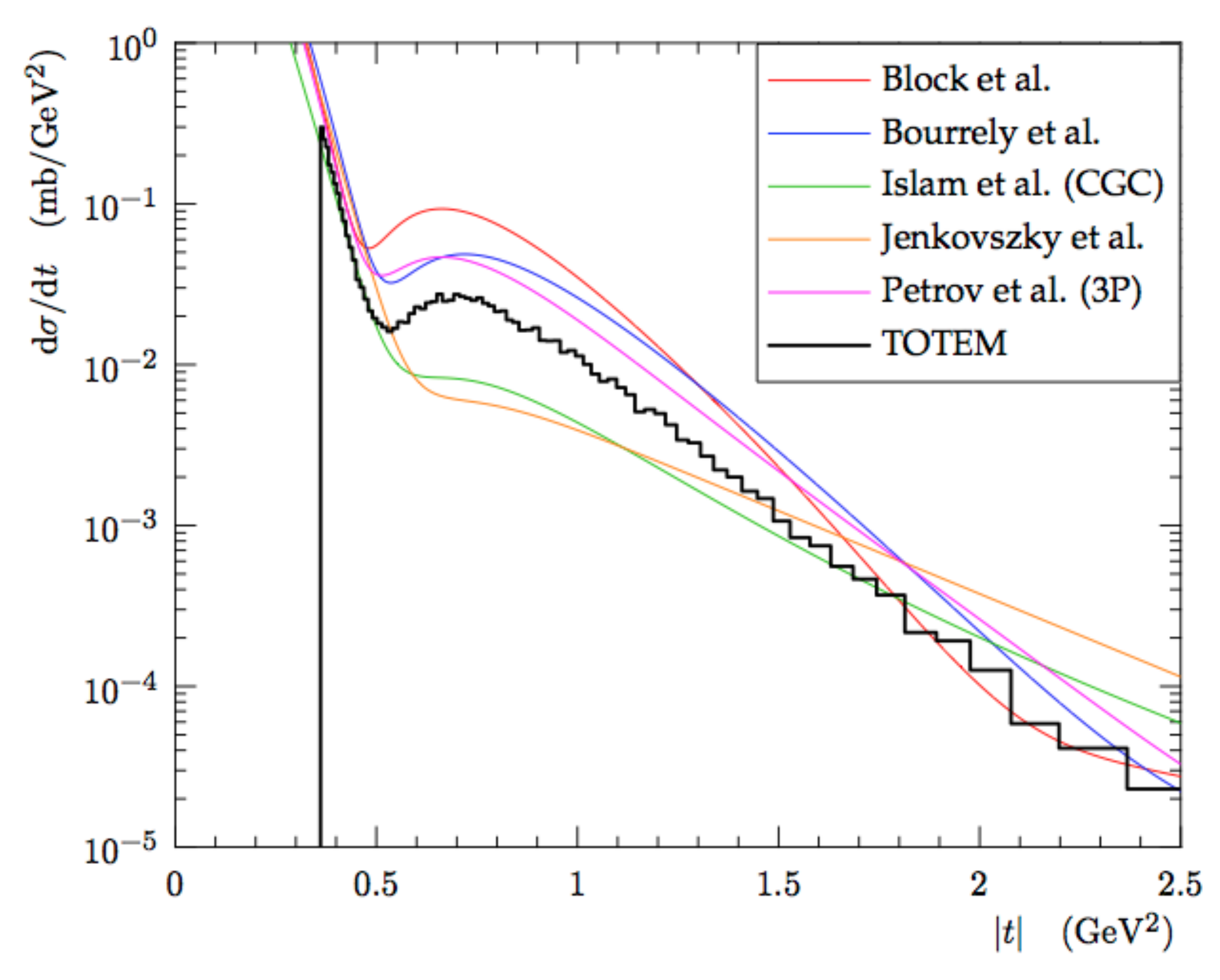}}
\end{center}
\caption{$d\sigma/dt$ Elastic differential $pp$ cross-section for $t < 2.5 $ GeV$^2$ as measured by the TOTEM collaboration (black histogram) together with the predictions of several proton models.}
\label{fig:trange}
\end{figure}

\noi The TOTEM collaboration has also measured the differential $pp$ distribution in an extended $t$ range. 
As outlined in Section III, this measurement is a unique tool to identify the proton macro regions.
 Figure~\ref{fig:trange} shows this result together with the prediction of several proton models which do not quite reproduce the data well. \\

\noi In the TOTEM analysis, the value of \stot is obtained using Equation~\ref{eq:opt} while  the value of the inelastic cross section \sinel is obtained by subtraction: 
$$\sigma_{Inel}(pp) = \sigma_{Tot}(pp) - \sigma_{El}(pp).$$ 
Using data collected both at $\sqrt{s} =$ 7 and 8 TeV, the TOTEM collaboration has produced a complete set of measurements, shown in Table~\ref{tab:totem}.

\begin{table}[h]
  \begin{center}
  \begin{tabular}{c c  c c } \hline
Measurement & $\sigma_{Tot}(pp)$ &   $\sigma_{El}(pp)$ &  $\sigma_{Inel}(pp)$ \\ \hline
 $\sqrt{s} = 7 $ TeV &   98.6 $\pm$ 2.8 mb  &  25.4 $\pm$ 1.1 mb & 73.1 $\pm$ 1.3 mb \\
 $\sqrt{s} = 8 $ TeV &  101.7 $\pm$ 2.9 mb &  27.1 $\pm$ 1.4 mb & 74.7 $\pm$ 1.7 mb\\ 
\hline
\end{tabular}
\caption{Values of the total, elastic and inelastic $pp$ cross section at $\sqrt{s} = $ 7 and 8 TeV as measured by the TOTEM collaboration}
\label{tab:totem}
\end{center}
\end{table}

\section{Measurements of parts of   $\sigma_{Inel}(pp)$ and extrapolation to its total value.}

\noindent As outlined above, the complete measurement of all processes that compose  $\sigma_{Inel}(pp)$ is very difficult and currently no experiment is able to do it. Both cosmic-rays and collider experiments can directly measure only parts of  $\sigma_{Inel}(pp)$ and therefore need to use MCs or analytical models to provide an estimate of the full values of $\sigma_{Inel}(pp)$. \\

\noi The energy range of collider and cosmic-ray experiments is largely overlapping, however the highest energy is reached only by the AUGER and HiRes experiments.  The cosmic-rays EAS-TOP experiment  overlaps in energy with  the Tevatron and LHC low energy runs (1-2 TeV) while the cosmic-rays experiments AGASA and Fly's Eye overlap with the high energy runs of LHC (7-14 TeV). The cosmic-rays experiment HiRes and AUGER measures $\sigma_{inel}$ in the energy range 50-100 TeV, well above current collider experiments, Figure~\ref{fig:CR_Comp}.

\section{Cosmic-rays experiments}
\noindent Cosmic-rays experiments measure the interaction of the primary particle with a nucleus in the atmosphere via the detection of secondary particles, generated in the hadronic shower, that reach ground level. This method is bound to carry significant uncertainties as the measured quantities are only indirectly related to the primary scattering event: the cosmic-rays flux composition, the atmospheric molecular mixture, the modelling of the hadronic shower, and the limited detector acceptance concur to the measurement uncertainties.  \\

\begin{figure}[h]
\begin{center}
 \resizebox{7cm}{!}{\includegraphics{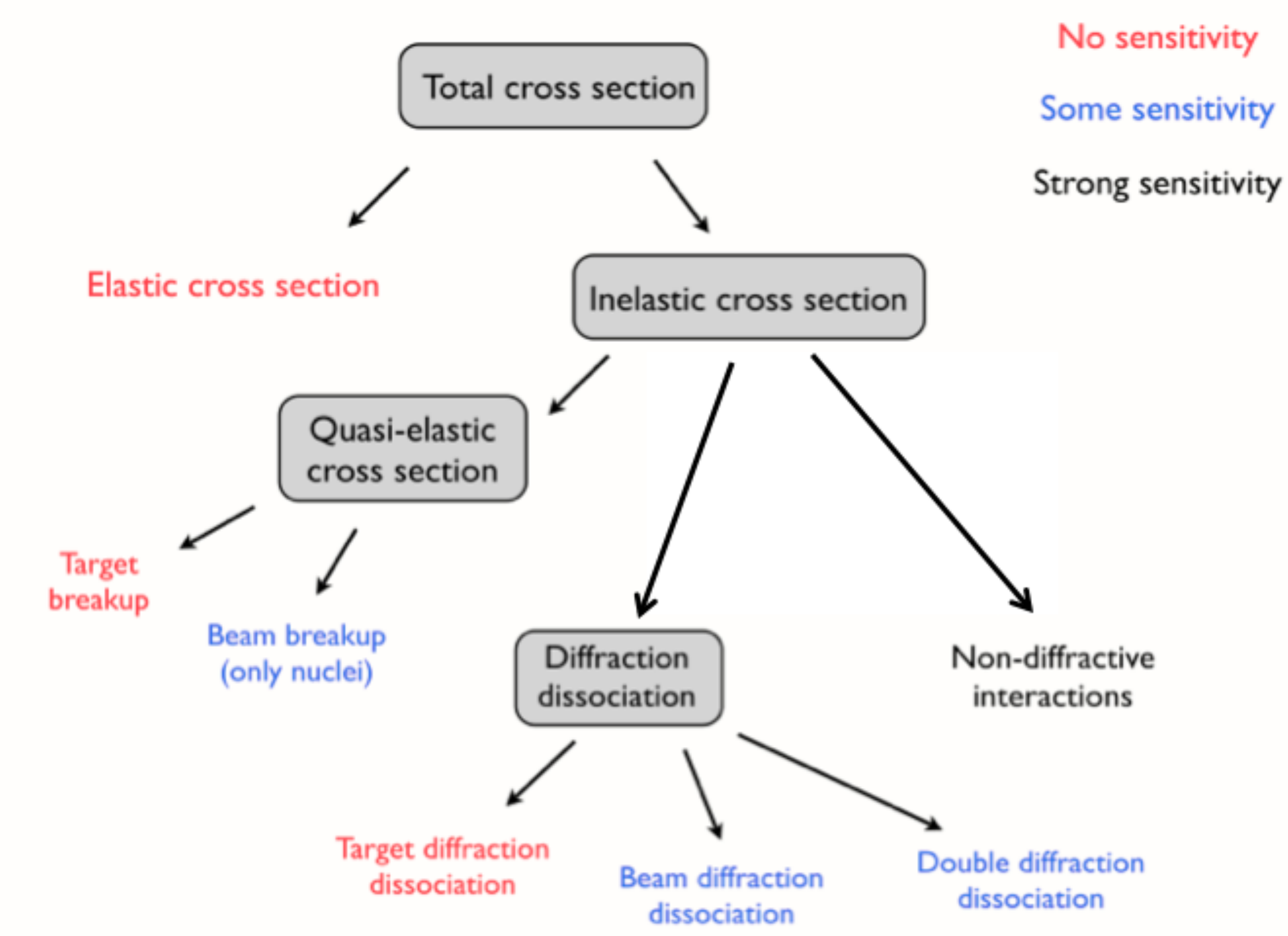}}
\end{center}
\caption{Sensitivity of the  cosmic-ray $p-air$ cross-section measurement to the various parts  of the total cross section (adapted from~\cite{Ulrich:2009zq}). }
 \label{fig:CR_parts}
\end{figure}

\noi The measurement is sensitive mostly to the non-diffractive part of the inelastic cross section,  it has moderate sensitivity to diffractive dissociative processes of the incoming primary particle and no sensitivity to other processes, Figure~\ref{fig:CR_parts}. 

\begin{figure}[h]
\begin{center}
 \resizebox{6cm}{!}{\includegraphics{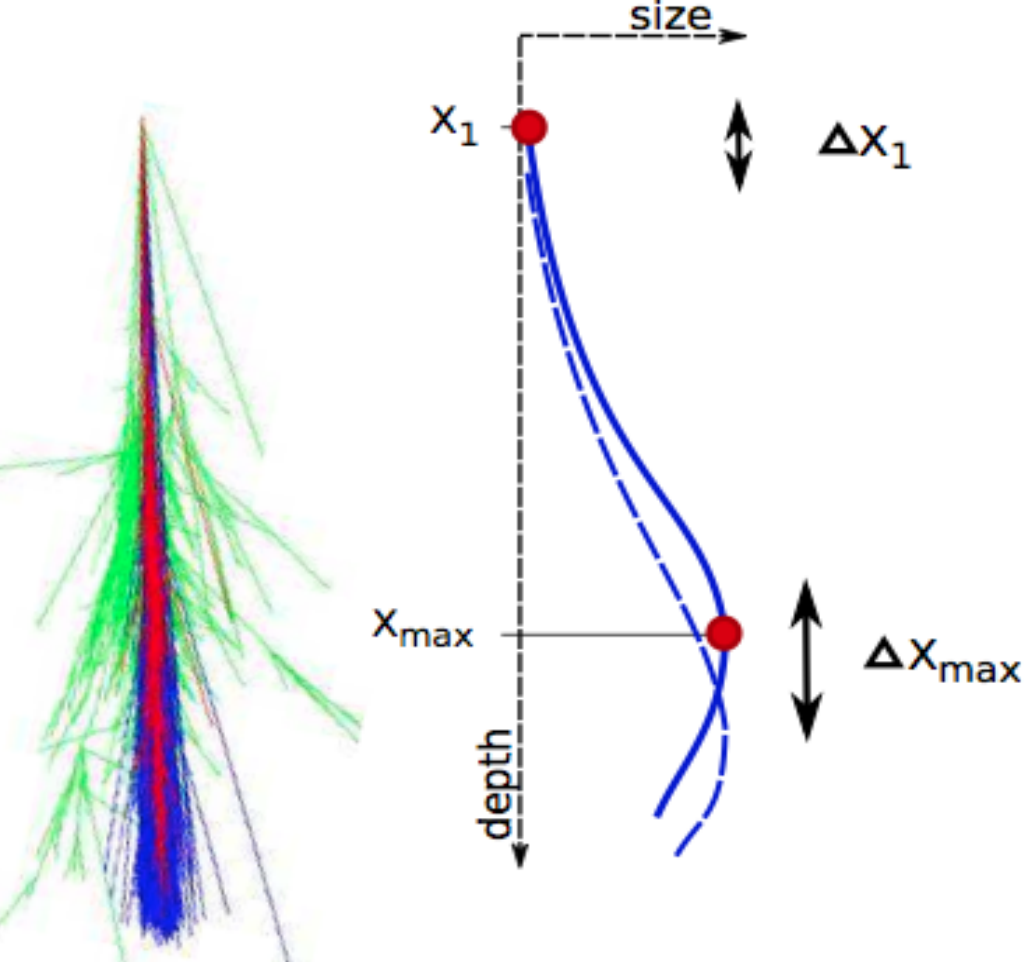}}
\end{center}
\caption{Sketch of a cosmic-ray shower development. $x_1$ indicates the position of impact. }
 \label{fig:cos2}
\end{figure}

\noi The value of $\sigma_{Inel}(p-air)$  directly influence  the distance $x$ that the primary particle travels in air before interacting, Figure~\ref{fig:cos2} : lower values of the cross section move  the point of interaction $x_1$ deeper into the atmosphere.   There are two main methods to reconstruct the $x_1$ position. 

\begin{figure}[h]
\begin{center}
 \resizebox{6cm}{!}{\includegraphics{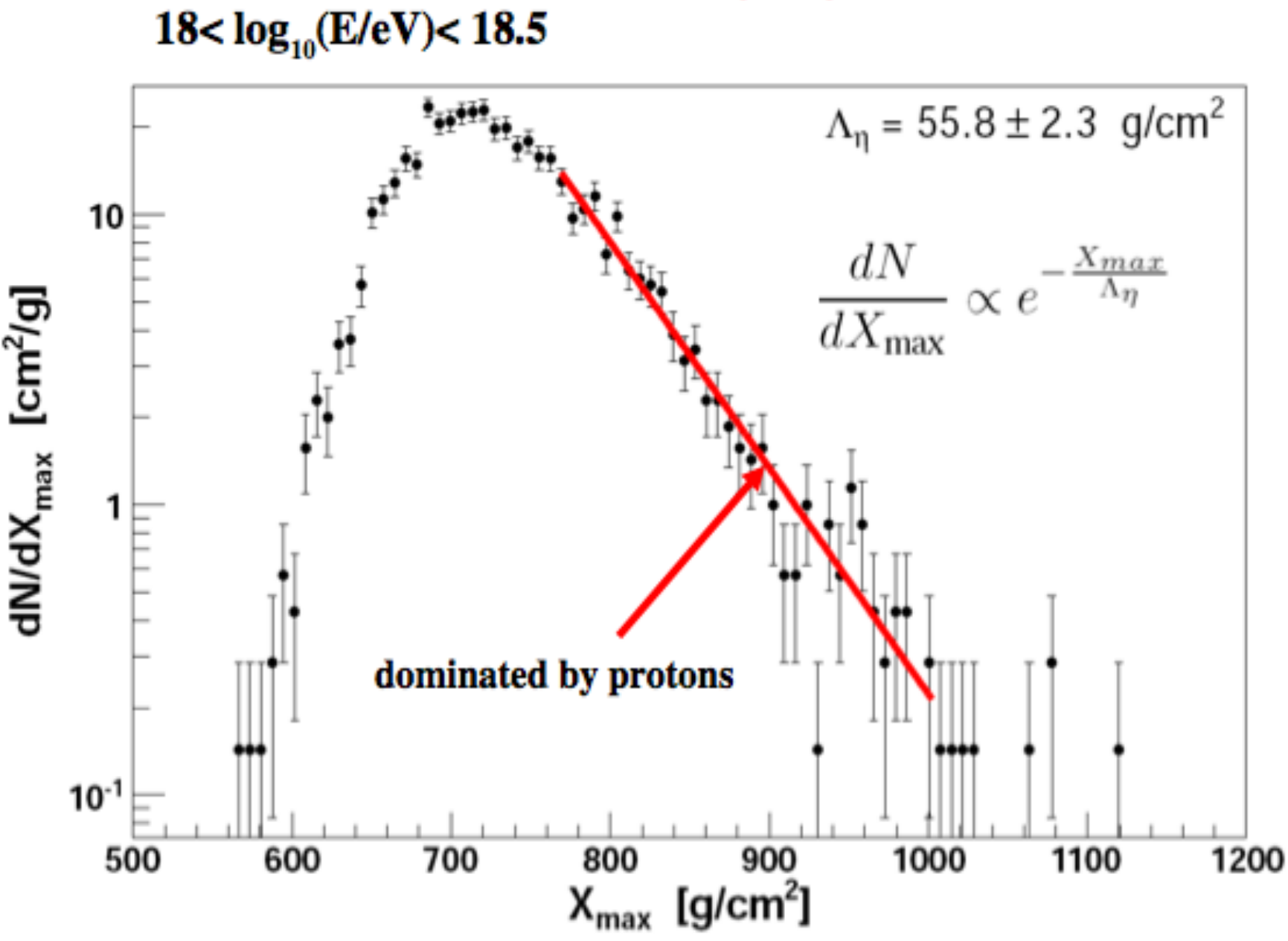}}
\end{center}
\caption{Position of shower maximum as a function of depth in the atmosphere as measured by the AUGER collaboration. }
 \label{fig:cos12}
\end{figure}

\begin{itemize}
\item[(i)] $\frac{N_e}{N_\mu}$: the ratio of the number of electrons to the number of muons is related to the shower length. By measuring this ratio together with the shower direction, the position of $x_1$ can be determined. This method relies on  MC models to simulate the shower development and to correctly predict the ratio $\frac{N_e}{N_\mu}$ as a function of shower depth. 
\item[(ii)] $X_{Max}$-tail: for fixed energy of the primary particle,  the probability of having a shower maximum deeper and deeper in the atmosphere decreases exponentially and therefore fitting the distribution of the position of shower maximum as a function of the depth allows to reconstruct $x_1$, Figure~\ref{fig:cos12}. To  select a sample of primary particles rich in protons, the  deeper tail of the distribution is used in the fit as heavier primary particles interact earlier on. 
\end{itemize}

\begin{figure}[h]
  \begin{center}
\resizebox{8cm}{!}{\includegraphics{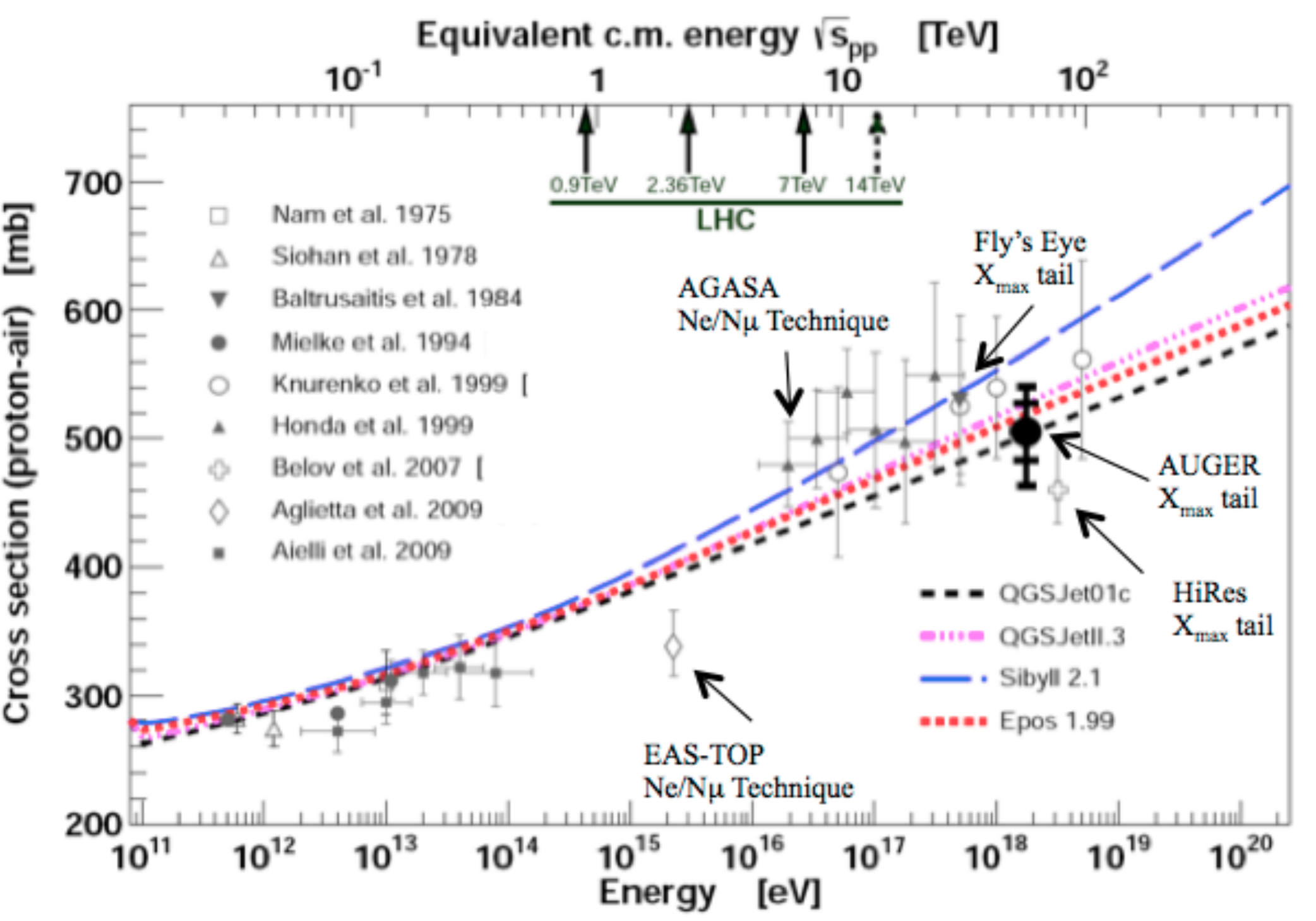}}
  \end{center}
    \caption{Values of the $p-air$  inelastic total cross section as a function of  \rs}
     \label{fig:CR_Comp}
\end{figure}

\noi The AUGER collaboration~\cite{Collaboration:2012wt}, using this second method, has recently published a new result for the proton-air inelastic cross section at 57 TeV: 
$$ \sigma^{57 TeV}_{Inel} (p-air) = 505\; \pm \; 22\; (stat) ^{+ 28} _{-36}\; (syst) \; mb.$$ 
Figure~\ref{fig:CR_Comp} shows a compilation of the results of $p-air$ total inelastic cross section (adapted from~\cite{Ulrich:2009zq}).

\noi The range of values used in the fit   is chosen so that the  remaining helium and heavy nuclei contribute less than the statistical uncertainty. The first step of the measurement is the evaluation of $\Lambda_\eta$ which, using MC models, is linked to the value of \sinelair. The determination of  $\Lambda_\eta$ has a systematic uncertainties of $\pm 15$ mb, while the remaining part of the systematic uncertainty is due to the process of extracting  \sinelair. \\

\begin{figure}[h]
  \begin{center}
\resizebox{8cm}{!}{\includegraphics{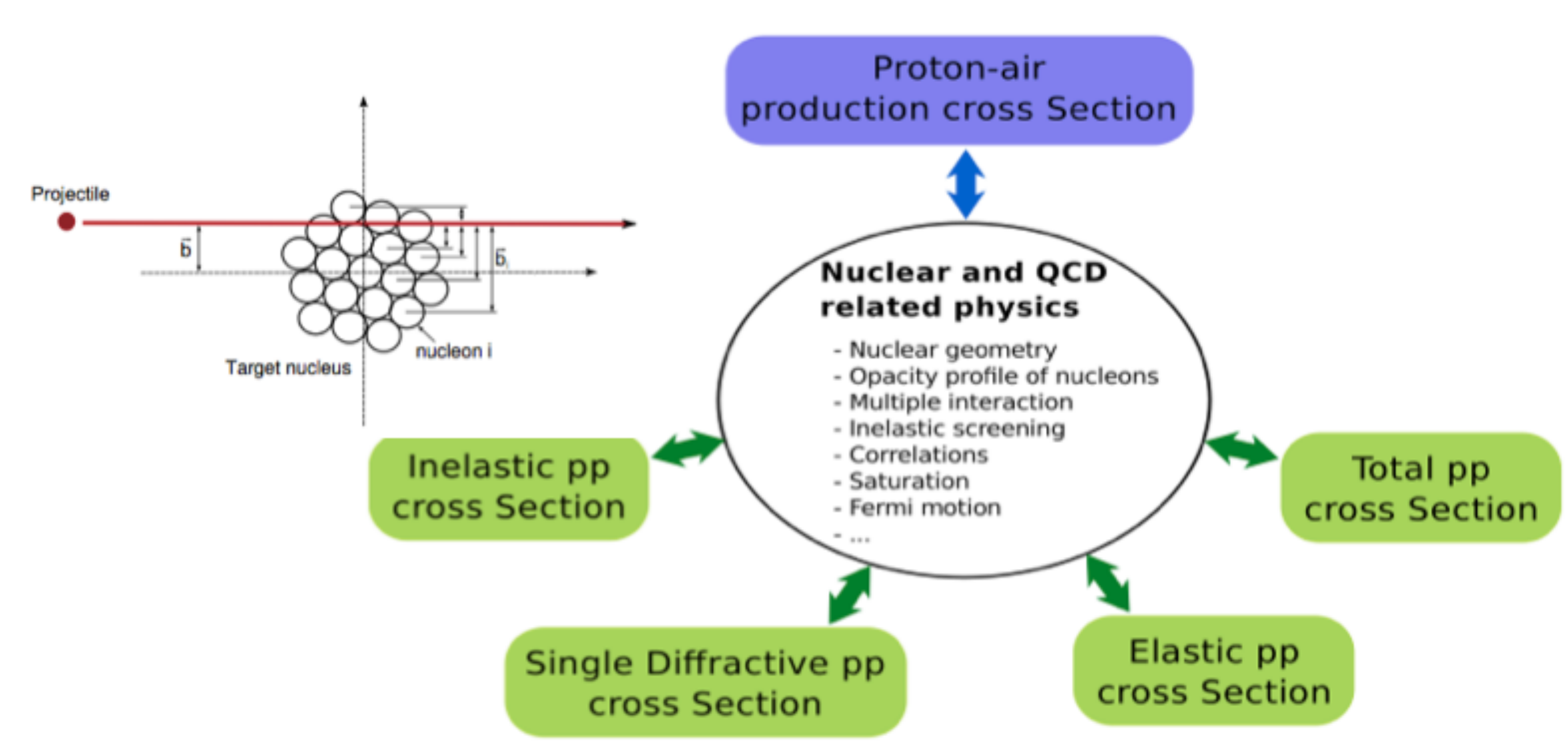}}
  \end{center}
    \caption{ Schematic of the Glauber components used to correlate the values of the $p-air$ and $pp$ cross sections.}
     \label{fig:glauber}
\end{figure}

\noi From \sinelair, the values of the proton-proton inelastic and total cross-sections can be obtained using the Glauber model (for an introduction see~\cite{Shukla:2001mb}) that  describes the proton-nucleus (and also nucleus-nucleus) scattering as a sum of elementary nucleon-nucleon interactions. The Glauber model takes into account various nuclear and QCD effects such as nuclear geometry, opacity of nucleons, multiple interactions, diffraction, saturation, Fermi motion, the total, inelastic and elastic cross sections at lower energies and the value of \sinelair at 57 TeV to provide an estimate of the proton-proton total and inelastic cross sections, Figure~\ref{fig:glauber}.

\noi The AUGER collaboration finds that in the Glauber framework the inelastic cross-section is less dependent on model assumptions than the total cross-section. The result for the inelastic  and total proton-proton cross-sections are 
$$\sigma^{57 TeV}_{Inel} (pp) = 92\; \pm \; 7\; (stat) \pm 9 \; (syst) \pm 7 \;\; (Gl.)  \;\; mb $$ 
and 
$$\sigma^{57 TeV}_{Tot} (pp) = 133\; \pm \; 13\; (stat) \pm 17 \; (syst) \pm 16 \;\; (Gl.) \;\; mb.$$

\section{Collider experiments}
\noi Collider experiments such as ALICE, ATLAS and CMS are able to directly measure the fraction of $\sigma_{Inel}(pp)$ composed by those events that leave {\it enough} energy in the  detector, typically in a rapidity interval $\eta \le |5|$, where the definition of {\it enough} is experiment and technique dependent. The results published so far rely on two different methods: (i) Pile-up counting (ii) Forward energy deposition.

\subsection{Pile-up method to determine $\sigma_{Inel}$}

\noi This method, used for the first time by the CMS collaboration~\cite{Chatrchyan:2012nj}, assumes that the number $n$ of inelastic $pp$ interactions in a given bunch crossing follows the Poisson probability distribution:

\begin{equation}
\label{eq:pois}
P(n,\lambda) = \frac{\lambda^n e^{-\lambda}}{n!},
\end{equation}
\noindent where $\lambda$ is calculated from the product of the instantaneous luminosity for a bunch crossing and the total inelastic $pp$ cross section: $\lambda = L * \sigma_{Inel}(pp)$. This technique counts the number of vertices in different luminosity intervals, and performs a fit using equation~\ref{eq:pois}. By construction, this method is sensitive only to those events that produce a detectable vertex. To collect an unbiased sample of $pp$ interactions, the event selection is performed using a  high-$p_t$ muon trigger, which is completely  insensitive to the presence of pile-up. Note that the event containing the triggering muon is not counted towards the total number of vertices for a given $pp$ crossing.\\

\begin{figure}[h]
\begin{center}
 \resizebox{9cm}{!}{\includegraphics{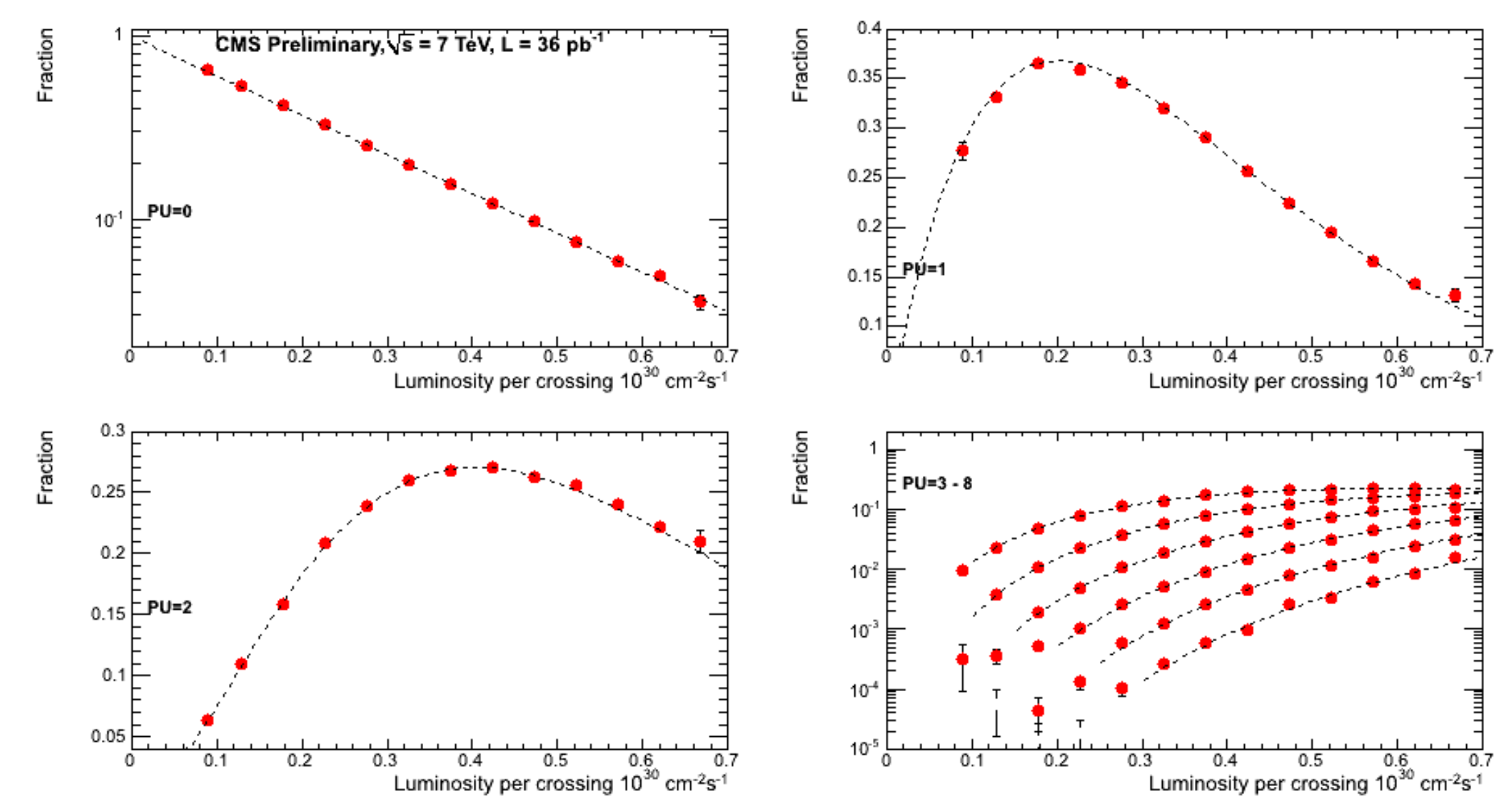}}
\end{center}
\caption{Fraction of $pp$ events with $n$ pileup vertices, for $n$ = 0 to 8, containing more than one track as a function of instantaneous bunch-crossing luminosity. The dashed lines are the fits using Equation~\ref{eq:pois}. }
\label{fig:pois}
\end{figure}

\noi  The probability of having $n$ inelastic $pp$ interactions ($n$ between 0 and 8), each producing a vertex with at least two  charged particles with $p_\perp > $ 200 MeV/c within $|\eta|$ = 2.4,  is measured at different luminosities to evaluate $\sigma_{Inel}(pp)$ from a fit of Equation~\ref{eq:pois} to the data, Figure~\ref{fig:pois}.  For each $n$, the values of the Poisson distribution given by Equation~\ref{eq:pois} are fitted as a function of $\lambda = L * \sigma_{Inel}$ to the data, providing an estimates of $\sigma_{Inel} (pp)$.

\subsection{Forward energy method to determine $\sigma_{Inel}$}

\noi The basic idea of this method is to count the number of events that, in a given interval of  integrated luminosity, deposit  at least a minimum amount of energy  in either of the forward parts of the detector; the number of events is then converted into a value of cross section by accounting for detector  and pile-up effects.  ATLAS and CMS require at least $E_{min} = 5$ GeV in the rapidity interval $3 \le |\eta| \le 5$ which is equivalent to, according to MC studies,  a minimum hadronic mass  $M_X$ of  at least 16  GeV/c$^2$ ( $\xi = M_X^2/s > 5 * 10^{-6}$). \\

\noi The TOTEM collaboration, using their T1 and T2 forward detectors ( 3.1 $\le |\eta| \le $ 6.5),  measured \sinel for events with at least one $|\eta| \le 6.5$ final-state particle~\cite{Antchev:2013bb}. This measurement includes all events with $M_X > 3.4 $ GeV, aside for a small component of central diffractive events of which maximally 0.25 mb is estimated to escape the detection of the telescopes.  \\

\noi The two techniques outlined above  are complementary:  the {\it pile-up} technique  measures more central events, while the {\it forward energy} technique is geared towards more forward topologies. Their combination  covers almost completely all type of events with a minimum particle production in the pseudorapidity interval $-6.5 < \eta < 6.5$. The comparison of these results with various MCs provides interesting information to model builders, as they test complementary parts of particle production.

\subsection{Results on $\sigma_{Inel}(pp)$}

\begin{figure}[h]
\begin{center}
 \resizebox{8cm}{!}{\includegraphics{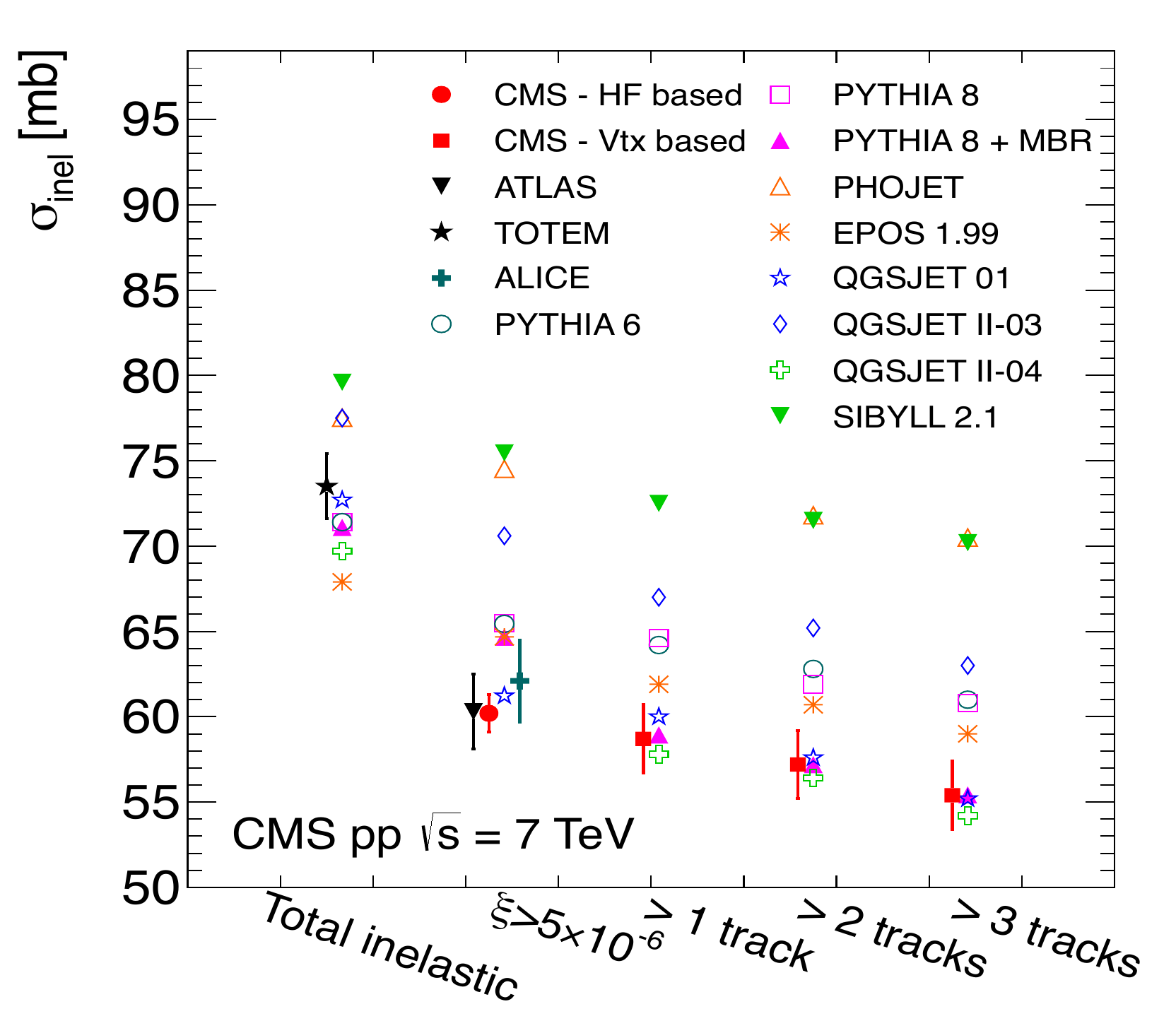}}
\end{center}
\caption{Compilation of the ALICE, ATLAS and CMS measurements of the inelastic $pp$ cross section  compared to predictions from several Monte Carlo models for different selection criteria, as labelled below the abscissa axis. The MC predictions have an uncertainty of 1 mb.}
\label{fig:res}
\end{figure}

\noindent Table~\ref{tab:all} lists the results from the ALICE~\cite{Abelev:2012sea}  ATLAS~\cite{Aad:2011eu}, CMS~\cite{Chatrchyan:2012nj} and  TOTEM~\cite{Antchev:2013bb} Collaborations for several selection criteria. The ALICE, ATLAS and TOTEM collaborations have extrapolated the measured values to provide also an estimate of $\sigma_{Inel}^{Tot}(pp)$. As the TOTEM collaboration has the most forward pseudorapidity reach, their \sinel value has the smallest systematic error. Figure~\ref{fig:res} shows a compilation of the  results for different selection criteria, and a comparison with several MCs predictions.

\begin{table}[h]
  \begin{center}
  \begin{tabular}{c c  c c c c } \hline
Exp &  Measurement  &  Result &  Stat & Syst & Lum \\ \hline
ALICE & $\sigma_{Inel}^{(\xi > 5\times 10^{-6})}$ &    62.1 &   & $  ^{+1.0}_{-0.9} $ & $ \pm 2.2$  mb \\
ATLAS & $\sigma_{Inel}^{(\xi > 5\times 10^{-6})}$ &    60.3 &$ \pm 0.05 $ & $ \pm 0.5 $ & $ \pm 2.1$ mb \\
CMS &$\sigma_{Inel}^{(\xi > 5\times 10^{-6})}$ &    60.2 & $ \pm 0.2 $&$  \pm 1.1 $&$ \pm 2.4 $ mb \\ 
TOTEM & $\sigma_{Inel}^{(\xi > 2.4\times 10^{-7})}$ &    70.5 & $\pm 0.1 $ &$  \pm 0.8 $&$ \pm 2.8 $ mb \\ \hline
ALICE & $\sigma_{Inel}$ &    73.2 &   & $  ^{+2.0}_{-4.6} $ & $ \pm 2.6$  mb \\
ATLAS & $\sigma_{Inel}$ &    69.4 &  & $ \pm 6.9 $ & $ \pm 2.4$ mb \\ 
TOTEM & $\sigma_{Inel}$ &    73.7 & $\pm 0.1 $ &$  \pm 1.7 $&$ \pm 2.9 $ mb \\ \hline
CMS &  $\sigma_{Inel}^{({>}1\mbox{ track})}$ &  58.7 & & $ \pm 2.0 $&$ \pm 2.4 $ mb  \\
CMS &  $\sigma_{Inel}^{({>}2\mbox{ tracks})}$ & 57.2 & & $  \pm 2.0 $&$\pm 2.4 $ mb   \\
CMS &  $\sigma_{Inel}^{({>}3\mbox{ tracks})}$ &  55.4 & & $ \pm 2.0 $&$ \pm 2.4 $mb   \\ \hline
\hline
\end{tabular}
\caption{$\sigma_{inel}(pp)$ values for $\xi >  5\times 10^{-6}$, $\xi >  2.4\times 10^{-7}$  and for interactions with $>$1, $>$2 and $>$3 charged particles in the final state, with their uncertainties from systematic sources of the method and from luminosity.}
\label{tab:all}
\end{center}
\end{table}

\noi The data points are compared to a large set of predictions from many MC models, used both in cosmic-rays physics and collider experiments. Although several Monte Carlo models such as EPOS, QGSJET 01, QGSJET II-4, PYTHIA 6, and PYTHIA 8 reproduce correctly the value of $\sigma_{Tot}(pp)$, only QGSJET 01, QGSJET II-04, and PYTHIA 8-MBR (but less	so)	are	able	to	simultaneously	reproduce	the	less	inclusive measurements. This observation suggests that most of the Monte Carlo models overestimate the contribution from high-mass events to the total inelastic cross section, and underestimate the component at low mass.

\section{Measurement of the diffractive component of $\sigma(pp)$}

\noi The most used technique to select diffractive events is the request of a gap in the rapidity distribution of final state particles. This method, however, has several drawbacks: rapidity gaps can also appear in non-diffractive events, albeit less frequently and high mass diffractive events often have a rapidity gap that is too small to be used as a signature and therefore they cannot be  selected using this method.

\begin{figure}[h]
\begin{center}
 \resizebox{6cm}{!}{\includegraphics{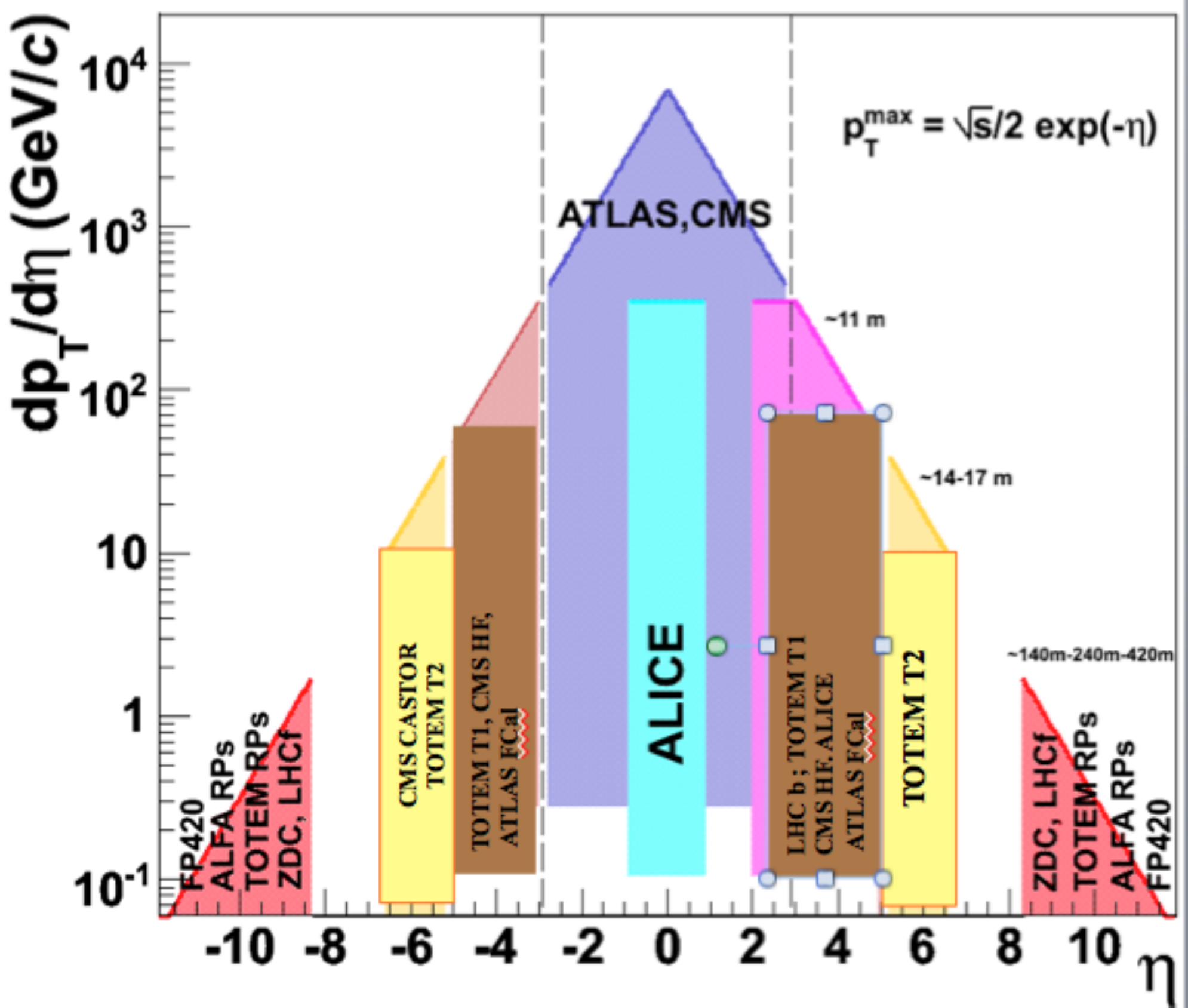}}
\end{center}
%\caption{Left pane: Rapidity - $p_\perp$ coverage of the LHC experiments at $\sqrt(s) = 14$ TeV. Right pane: sketch of the rapidity coverage and rapidity gap of an event with mass $M_X$}
\caption{Rapidity - $p_\perp$ coverage of the LHC experiments at $\sqrt(s) = 14$ TeV.}
 \label{fig:coverage}
\end{figure}

\noindent   At LHC, the rapidity interval is roughly 20 units, while the experimental coverage of the various experiments is much less, Figure~\ref{fig:coverage}: ATLAS and CMS cover roughly 10 units of rapidity, centrally from -5 to 5, TOTEM covers only forward rapidities,  3.1 $< |\eta |< 6.5$ while ALICE has an asymmetric coverage, $ -3.4 < \eta<  5.1 $. \\

\begin{figure}[h]
\begin{center}
 \resizebox{7cm}{!}{\includegraphics{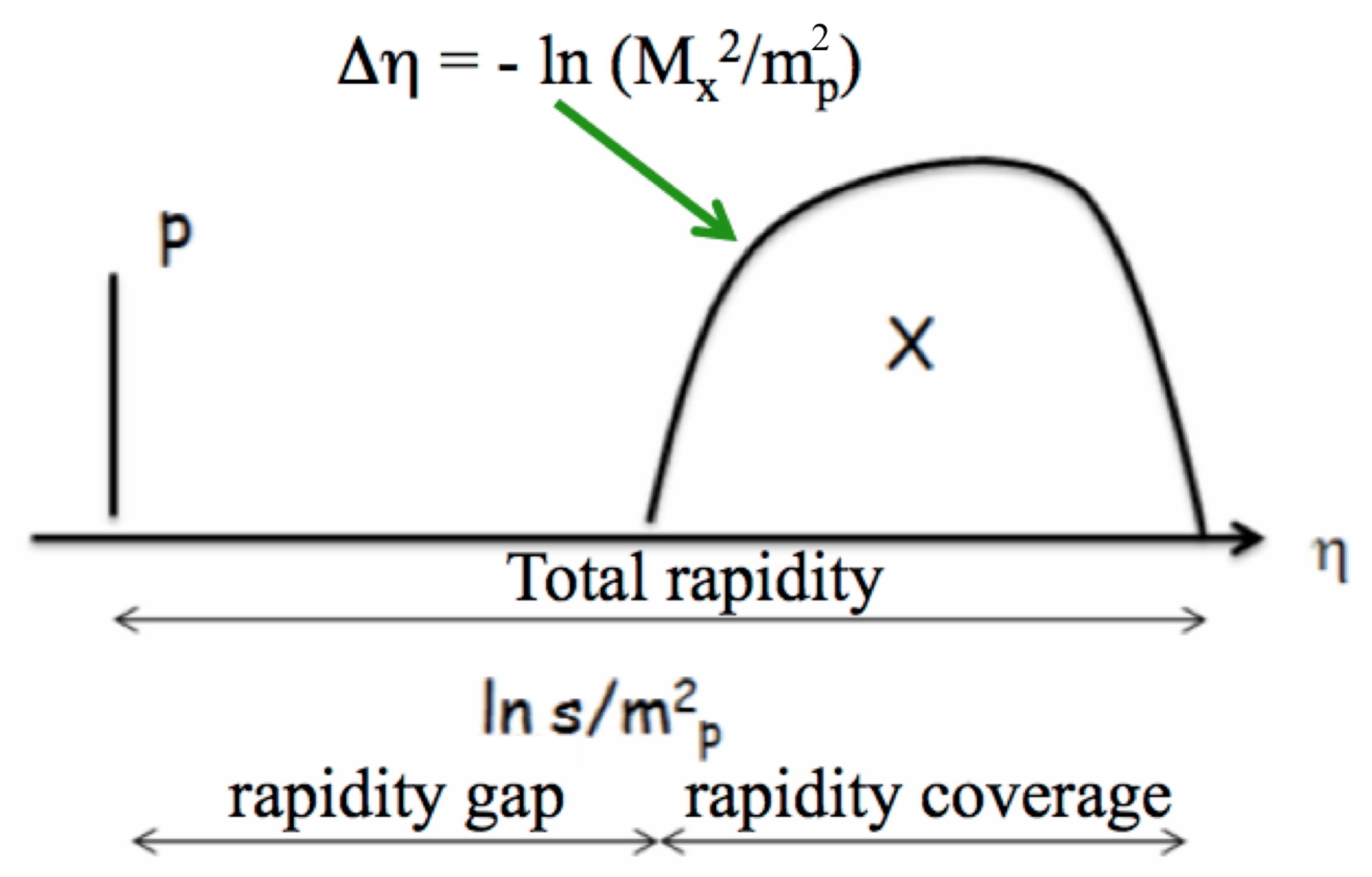}}
\end{center}
\caption{Sketch of the rapidity coverage and rapidity gap of an event with mass $M_X$}
 \label{fig:mx}
\end{figure}

\noi The rapidity span of an event with mass $M_X$ is given by $\Delta\eta = ln (M^2_X /m^2_p ) $ while the rapidity gap of the event is given by $\Delta\eta = -ln \xi$, with $m_p$ the proton mass and $\xi = M^2_X/s$, Figure~\ref{fig:mx}. For this reason small mass events are very difficult to measure as  they are boosted forward and don't leave in the detector any signature: an event with $M_X = $ 5 GeV covers only $\sim$ 3 unit of rapidity and therefore escapes detection.  The experimental limit of detection for both ATLAS and CMS is $\xi > 5 * 10^{-6}$, which correspond to a mass $M_X \simeq 15$ GeV at 7 TeV while TOTEM reaches down to  $M_X \simeq 3.5$ GeV. \\

\noi The selection of diffractive events  is defined by each experiment differently, relying on the strengths of its own detector: ATLAS requires 4 units of rapidity gap for {\it single diffractive } events and 3 units for    {\it double diffractive } events within the detector acceptance, TOTEM uses separately the information on T1 and T2 to select low and high mass events while CMS requires no activity with either $\eta <1$ or $\eta>-1$. Several experiments have measured the cross section values of \ssd for different intervals of the hadronic mass, Table~\ref{tab:sd}. ATLAS has reported the fraction of $pp$ events with  a rapidity gap in the interval $2.09 < \eta < 3.84$ to be $f_{GAP}$ = 10\% ~\cite{Aad:2011eu}.

\begin{table}[h]
  \begin{center}
  \begin{tabular}{c c  c c} \hline
Experiment & Energy & Mass &   \ssd\\ \hline
& [TeV] & [GeV] &   [mb]\\ \hline
TOTEM  & 7 & 3.4 -  1100 & 6.5 $\pm$ 1.3 \\
(preliminary) & & & \\
CMS  & 7 & 12 -  394 & 4.27 $\pm$ 0.04 (sta) $^{+0.65}_{-0.58}$ (sys) \\
ALICE & 2.76 & 0 - 200  & 12.2 $ \pm ^{+3.9}_{-5.3}$ \\ 
ALICE & 7 &  0 - 200 & 14.9 $\pm ^{+3.4}_{-5.9}$ \\ 
\hline
\end{tabular}
\caption{Values of the single diffractive \ssd cross section  as measured by TOTEM~\cite{DIS2013}, CMS~\cite{DIS2013}, ALICE~\cite{Abelev:2012sea}}
\label{tab:sd}
\end{center}
\end{table}

\section{Discussion of the results}
\begin{figure}[h]
%\vspace{-9cm}
\begin{center}
 \resizebox{8cm}{!}{\includegraphics{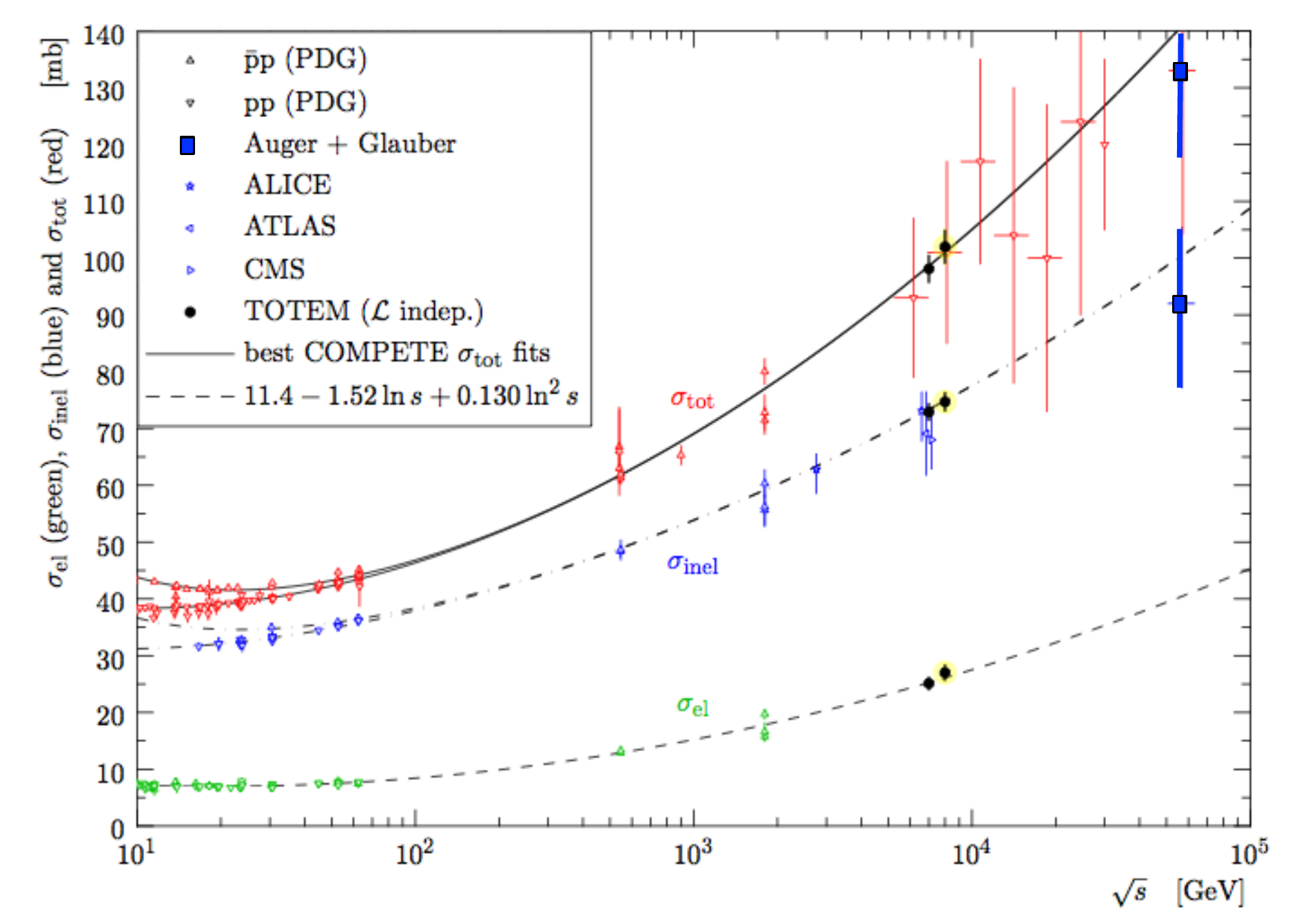}}
\end{center}
\caption{Compilation of the values of $\sigma_{Tot}(pp), \; \sigma_{El}(pp)$ and $\sigma_{Inel}(pp)$  as a function of \rs. The best fits from the COMPETE collaboration are also shown.}
\label{fig:total}
\end{figure}

\noi  Figure~\ref{fig:total} shows  a compilation of the values of $\sigma_{Tot}(pp), \; \sigma_{El}(pp)$ and $\sigma_{Inel}(pp)$  as a function of \rs.  The plot includes the results obtained by the TOTEM collaboration, together with the results from LHC, lower energy experiments and the best fit from the COMPETE collaboration based on a  $ln^2(s)$ behaviour of the cross section. 

\noi Several groups, for example  GLM~\cite{Gotsman:2010nw}~\cite{Gotsman:2012rq} (E. Gotsman, E. Levin and U. Maor), Durham~\cite{Ryskin:2011qe}\cite{Martin:2011gi}(M. G. Ryskin, A. D. Martin and V. A. Khoze), Ostapchenko~\cite{Ostapchenko:2010vb}, KP~\cite{Kaidalov:2009aw}(A. B. Kaidalov and M. G. Poghosyan) and DL~\cite{Donnachie:2011aa})(A. Donnachie and P. V. Landshof), that have proposed in the past models for soft interactions based on Pomeron exchanges have now included in their analyses the LHC results.  Some of them, such as Ostapchenko and DL, propose to use a soft and a hard pomeron, while other groups (GLM, Durham, KP) are using a single pomeron. A summary of the results is shown in Tab~\ref{tab:mod} (based on~\cite{Gotsman:lishep}). 

\begin{table}[h]
  \begin{center}
  \begin{tabular}{c c  c c c c } \hline
                     & DL &   Ost. &  GLM  & Durham & KP \\ \hline \hline
$\Delta_{soft}$       & 0.09 & 0.14  & & & \\
$\alpha\prime_{soft}$ & 0.25 & 0.14  & & & \\ \hline
$\Delta_{hard}$       & 0.36 & 0.31 & & & \\
$\alpha\prime_{hard}$ & 0.1 & 0.85 & & & \\ \hline 
$\Delta$             &  &  & 0.23 & 0.14 & 0.12 \\
$\alpha\prime$       &  &  & 0.028 & 0.1 & 0.22 \\
\hline
\end{tabular}
\caption{Values of the intercept and slope parameters of the pomeron trajectory in various models of soft interaction.}
\label{tab:mod}
\end{center}
\end{table}

\noi Bloch and Halzen~\cite{Block:2012ym}, updating their analyses to include the LHC results, have proposed a parametrization in the form of equation~\ref{eq:cross3} for both the total and inelastic cross section: 
\begin{eqnarray}
\sigma_{Tot} = 37.1s^{-0.5}+37.2-1.4ln(s)+0.3ln^2(s) \\
\sigma_{Inel}  =  62.6s^{-0.5}-0.5-1.6ln(s)+0.14ln^2(s) 
\end{eqnarray}
\noi It's worth stressing that the results of LHC had a strong impact on all models, showing that is quite difficult to extrapolate from lower to higher energy. \\

\noi  According to~\cite{Cudell:2009bx}, a value $\sigma_{Tot}^{\sqrt{s} = 14 TeV}(pp)$ = 120 - 160 mb is a clear sign of the two-pomeron model while  a value around 110 mb is an indication of the $ln^2(s) $ behaviour. \\

\subsection{Diffractive cross section, \ssd}
\begin{figure}[h]
%\vspace{-9cm}
\begin{center}
 \resizebox{8cm}{!}{\includegraphics{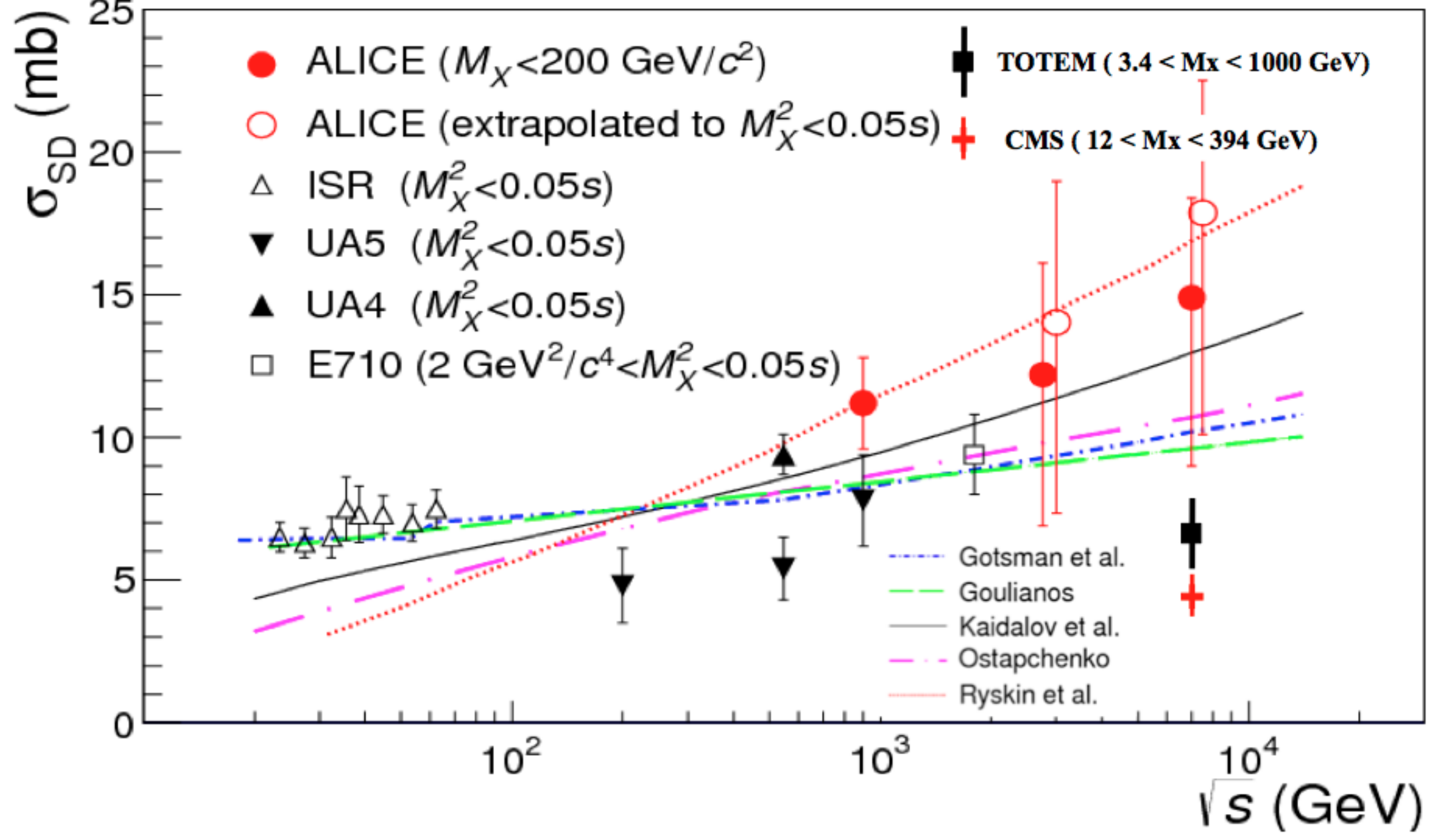}}
\end{center}
\caption{Compilation of  values of \ssd  as a function of \rs. Plot adapted from~\cite{Oe}.}
\label{fig:sd_plot}
\end{figure}

\noi Figure~\ref{fig:sd_plot} shows the experimental values and the predictions from some theoretical models of \ssd as a function of \rs.

\noi The experimental values of \ssd are listed in Table~\ref{tab:sd}.  The results from TOTEM and CMS have not been extrapolated outside the measured \mx range and therefore have smaller errors than the values reported by ALICE. An extrapolation of the CMS and TOTEM results to the  \mx range $M^2_X < 0.05 *s$ ($M_X \le 1550$ GeV) yields a value of \ssd $\sim$ 9-11 mb, compatible with several of the proposed models. It's interesting to note that even though the ALICE results seem to favour a higher value of \ssd, given the large experimental errors, all experimental points are compatible.\\

Table~\ref{tab:sdt} lists the predictions of several groups for the single and double diffractive $pp$ cross section values at $\sqrt{s} = 14 $ TeV.

\begin{table}[h]
  \begin{center}
  \begin{tabular}{c c  c c c c } \hline
                     & GLM1 & GLM2 & Durham (07) & Durham (11) & Ost \\ \hline \hline
\ssd       & 10.8 & 13.7  & 13.3 & 17.6-18.8& 11 \\
\sdd & 6.5 & 8.8  & 13.4 & 13.5 & 4.8 \\ \hline
\end{tabular}
\caption{Values of the single and double diffractive $pp$ cross sections  at $\sqrt{s}$ =  14 TeV in various models}
\label{tab:sdt}
\end{center}
\end{table}

\section{Comments on $\sigma(p\bar{p})_{Tot}$ at 1.8 TeV: CDF,  E710 and E811}
\noi Three experiments, E710~\cite{Amos:1989at}, E811~\cite{Avila:1998ej} and CDF~\cite{Abe:1993xy} , have measured the value of  $\sigma(p\bar{p})_{Tot}$ at 1.8 TeV obtaining the following results:

\begin{eqnarray}
{\rm E710:} \;\; \sigma(p\bar{p})_{Tot} = 72.8 \pm 3.1 \; {\rm mb} \\
{\rm E811:} \;\; \sigma(p\bar{p})_{Tot} = 71.4 \pm 2.4 \; {\rm mb} \\
{\rm CDF:}\;\; \sigma(p\bar{p})_{Tot} = 80.0 \pm 2.2 \; {\rm mb }
\end{eqnarray}

\noi In pre-LHC, pre-HERA era, the value of the cross section measured by E811 was considered the most reliable. This can be seen for example in Figure~\ref{fig:cross_sections} (top left pane) where the fit in the form of equation~\ref{eq:cross1} goes through the experimental point of E811 while passing well underneath of the CDF measurement. \\
\begin{figure}[h]
\begin{center}
 \resizebox{8cm}{!}{\includegraphics{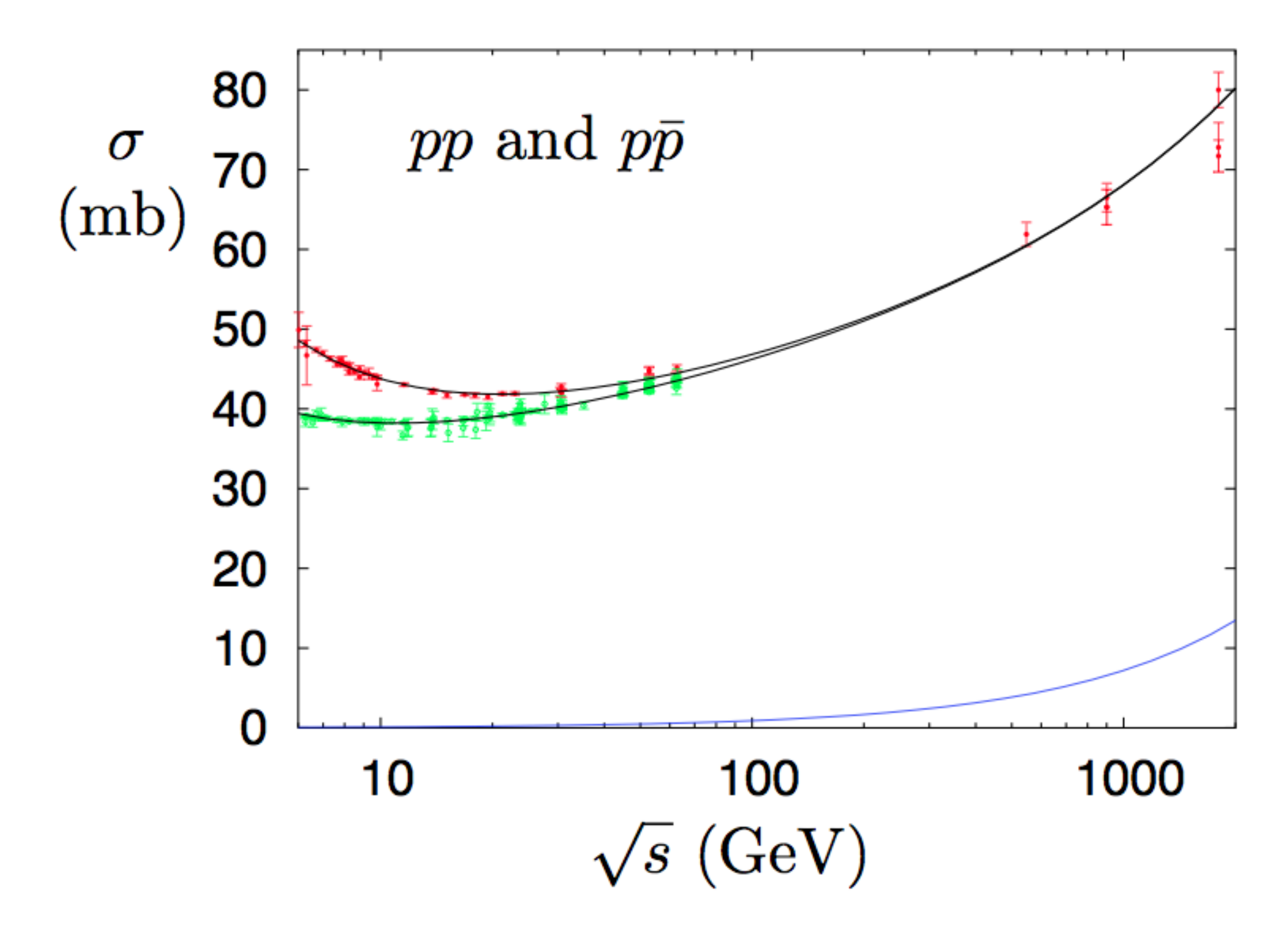}}
\end{center}
\caption{ Fit to the $pp$ and $p\bar{p}$ total cross section values using a two-pomeron parametrization. The bottom line show the contribution of the so called {\it hard pomeron} (\cite{Landshoff:2007uk}).}
\label{fig:postHera}
\end{figure}

\noi The introduction of the two-pomeron parametrization to accommodate the HERA data, shown in equation~\ref{eq:hard}, produced a  higher fit,  closer to the value measured by CDF,  Figure~\ref{fig:postHera}. \\

\noi The most recent fits of the COMPETE collaboration, see Figure~\ref{fig:total},  are also closer to the CDF value, indicating a growing consensus of considering this point more accurate.

\section{Conclusions}

\noi The concurrent efforts of several cosmic-rays and collider experiments have provided in the last couple of years a large quantity of measurements of the values of the total, elastic and inelastic cross sections as well as the values of cross sections for particular final state. The total value of the cross section is  well reproduced by the prediction of the COMPETE collaboration showing that a  $ln^2(s)$ dependence of $\sigma_{Tot}(pp)$ provide a good tool for extrapolating to higher energies. Several groups, using single or  double pomeron models, have updated their analyses using the new LHC data and also obtain good fits to the data. \\

\noi Common MC models used in collider experiments fail to concurrently reproduce the new measurements, pointing to an underestimation of the amount of low mass events.

\end{document}